\documentclass[10pt]{iopart}
\usepackage{graphicx}
\usepackage{iopams}

\begin{document}
\title{Fermi-to-Bose crossover in a trapped quasi-2D gas of fermionic atoms}
\author{A. V. Turlapov$^{1,2}$ and M. Yu. Kagan$^{3,4}$}
\address{$^1$Institute of Applied Physics, Russian Academy of Sciences, ul.~Ulyanova~46, 603000 Nizhniy Novgorod, Russia}
\address{$^2$N. I. Lobachevsky State University of Nizhni Novgorod, prosp.~Gagarina~23, 603950 Nizhniy Novgorod, Russia}
\address{$^3$P. L. Kapitza Institute for Physical Problems, Russian Academy of Sciences, ul.~Kosygina~2, 119334 Moscow, Russia}
\address{$^4$National Research University Higher School of Economics, ul.~Myasnitskaya~20, 101000 Moscow, Russia}
\eads{\mailto{turlapov$@$appl.sci-nnov.ru}, \mailto{kagan$@$kapitza.ras.ru}}
\vspace{10pt}
\begin{indented}\item[]\today\end{indented}
\begin{abstract}
Physics of many-body systems where particles are restricted to move in two spatial dimensions is challenging and even controversial: On one hand, neither long-range order nor Bose condensation may appear in infinite uniform 2D systems at finite temperature, on the other hand this does not prohibit superfluidity or superconductivity. Moreover, 2D superconductors, such as cuprates, are among the systems with highest critical temperatures. Ultracold atoms are a platform for studying 2D physics. Uniquely to other physical systems, quantum statistics may be completely changed in an ultracold gas: an atomic Fermi gas may be smoothly crossed over into a gas of Bose molecules (or dimers) by tuning interatomic interactions. We review recent experiments where such crossover has been demonstrated as well as critical phenomena in the Fermi-to-Bose crossover. We also present simple theoretical models describing the gas at different points of the crossover and compare the data to these and more advanced models.
\end{abstract}
\ioptwocol
\tableofcontents

\section{Introduction}

\subsection{The history of the Fermi-to-Bose crossover}

Change of quantum statistics, from bosonic to fermionic and back, may look trivial on one hand: Indeed, nearly all known bosons are composed of fermions, such as the hydrogen atom consists of a proton and electron, and, seemingly, the ionization resolves the problem of creating two Fermi systems from a Bose-condensate of hydrogen. On the other hand, the resulting Fermi systems would hardly be degenerate or stable.

For converting a degenerate bosonic system into a fermionic one and vice versa, smooth control of interactions is necessary. For the first time such problem was considered by Keldysh and Kozlov in 1968 \cite{ExcitonCrossover1968Eng} for a gas of excitons which substantially overlap at small electron-hole coupling and transform into a gas of point-like bosons as the coupling increases.

In the 1980s, it became clear that smooth transform of a Fermi gas into a Bose condensate is contained in the Bogolubov superconductivity theory~\cite{Bogoliubov1supercond1958eng,Bogoliubov3supercond1958eng} already. Even at infinitely small attraction, fermions with opposite spins form the Cooper pairs~\cite{CooperPairs1956,BCS1957,SchriefferBook1964}, while an adiabatic grow of attraction makes the pairs contract and smoothly cross over into point-like bosons~\cite{LeggettColloq1980,Leggett,Nozieres1985}. Such transformation of a fermionic system into a bosonic one is a fundamentally many-body effect. Initially the binding of the pairs appears due to presence of the Fermi surface. Only after the interaction strength overcomes a threshold, the bound state becomes stable in the vacuum.
The bound-state emergence is not accompanied by any jumps in system properties.
Such change in the statistics has been considered not only for the excitons but also for electrons in superconductors and for quark matter~\cite{QuarkCrossover2006}. Physical implementation of this phenomenon took place in experiment with an ultracold gas of Fermi atoms~\cite{Grimmbeta}.

\subsection{Two-body and many-body physics in 2D}

\begin{figure}[htb!]
\begin{center}
\includegraphics[width=\linewidth]{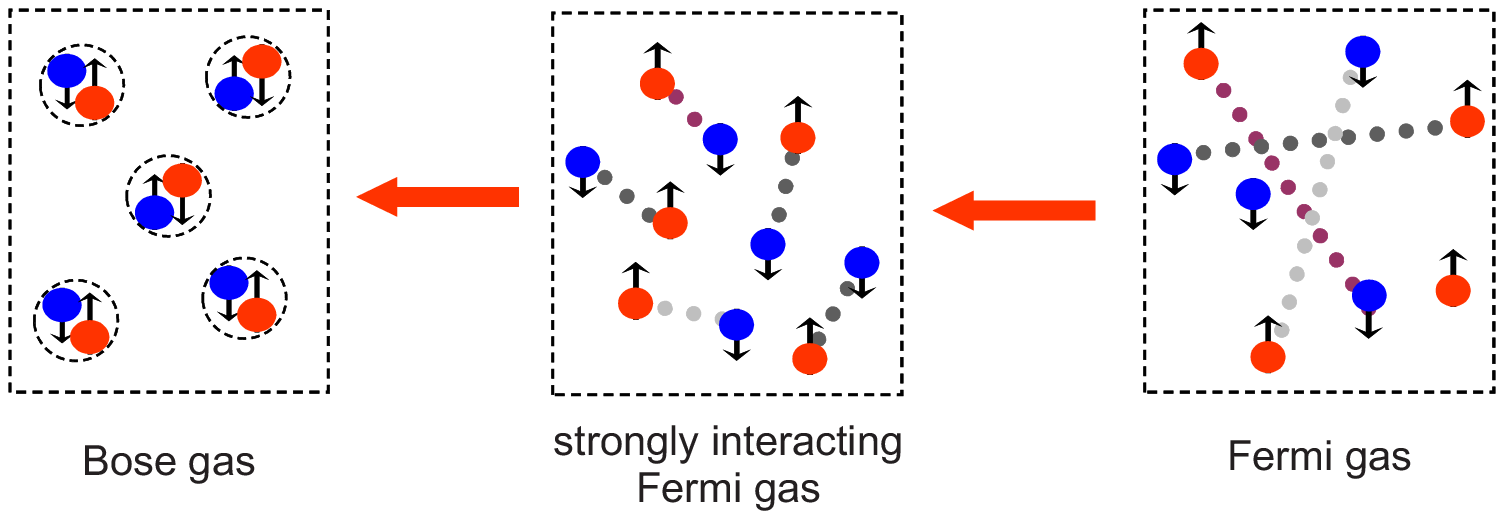}
\end{center}
\caption{Sizes of the pairs of fermions at different stages of the crossover between the Fermi (on the right) and Bose (on the left) statistics.}
\label{fig:Crossover}
\end{figure}

In a two-dimensional (2D) system, in case of attractive interaction of two particles, bound state always exists, even in vacuum: an infinitely small interaction may bind two fermions into a boson for a symmetric attractive potential. Nevertheless the problem of crossover between Bose and Fermi statistics naturally appears in a many-body two-dimensional system. Schematically such crossover is shown in figure~\ref{fig:Crossover}.
When the size of the vacuum bound state is larger than the interparticle distance, the system looks like a gas of fermions. Upon contraction of the bound state to the size much smaller that the interparticle distance, the system becomes bosonic. Between the fermionic and bosonic limits, there is the region of strong many-body interactions where the size of fermion pairs is comparable to the interparticle distance.

In uniform 2D systems the long-range order does not appear at finite temperature due to thermal fluctuations of the order parameter. In particular, this prohibits Bose condensation. Nevertheless the reduction of the spatial dimensionality from three to two does not impoverish many-body physics, but makes it more diverse. New phase transitions appear~\cite{BerezinskiiIIeng,KosterlitzThouless1973,VolovikTopoTrans1989,pTransition2005,Topo2DClassification2010}. For example, at the Berezinskii--Kosterlitz--Thouless (BKT) transition~\cite{BerezinskiiIIeng,KosterlitzThouless1973}, the superfluid phase is destroyed via proliferation of vortices.
The BKT transition became the first known phase transition with nontrivial topology. Presently topological phase transitions in 2D systems are actively studied~\cite{VolovikTopoTrans1989,pTransition2005,Topo2DClassification2010}. Other examples of interesting physics in 2D include the integer~\cite{IntegerHallReview2014eng} and fractional~\cite{FractionalQuantHallReview1999} quantum Hall effects. Explanation of high-temperature superconductivity observed in materials with two-dimensional kinematics is still an open question~\cite{KopaevEng}.

A unified theory of the Fermi-to-Bose crossover has to describe weakly-interacting Fermi and Bose gases as well as the gas in the intermediate strongly-interacting regime.
For 3D systems, generalization of the Bardeen--Cooper--Schrieffer (BCS) variational wavefunction (describing the paired state) has been successfully applied~\cite{LeggettColloq1980,Leggett,Nozieres1985}. Partially for this reason, the crossover problem is frequently referred to as the BCS--BEC crossover, where BEC stands for Bose--Einstein condensation.
For 2D, construction of such theory was more complicated than for 3D.
While in 3D a mean field theory of Cooper pairs~\cite{LeggettColloq1980,Leggett,Nozieres1985} qualitatively correctly describes ground-state properties, for a 2D system it is important to include the fluctuations into the model~\cite{SalasnichCompositeBosonsFromFluctuations2015,Fermi2DMeanFieldPlusFluctuations2015}. More important role of the fluctuations in lower dimensions is in agreement with the Ginzburg--Levanyuk criterion~\cite{GinzburgLevanyuk1959eng,GinzburgLevanyuk1960eng}. Note also that in recent years a considerable numerical effort has been applied to the problem of the BCS--BEC crossover in 2D in the framework of the advanced mean field~\cite{Fermi2DMeanFieldPlusFluctuations2015,HuiHuBKTviaFluctMF2017}, Monte-Carlo simulations~\cite{Giorgini2D2011,Fermi2DAbInitioLattice2015,Fermi2DExactGS2015,DiffusionMC2015}, and Luttinger--Ward (thermodynamic) identities~\cite{Fermi2DEOSandPressureParish2014}. Also a generalization of Nozi\`{e}res--Schmitt-Rink approach~\cite{Nozieres1985} with gaussian fluctuations to the 2D case was performed~\cite{SalasnichCompositeBosonsFromFluctuations2015,Salasnich2DFluct2016}.

The interaction description is a general question for kinematically 2D systems. In the theory of high-temperature superconductivity, for example, purely two-dimensional models are frequently used~\cite{LoktevReview2001}. These models assume that all motion is in the $xy$ plane and also the interactions are independent of $z$.
Such a pure two-dimensional model is not always sufficient, which may be seen in case of $^3$He on a substrate: The increase of zero-point oscillations relative to the interaction radius brings about a formation of a self-bound liquid state on the substrate of alkali metals (except for lithium)~\cite{Quasi2D3He2013}.
Unified and correct description of interactions in all regimes is required during construction of a model for a system with tunable statistics.
Experiments with such a system may test applicability of purely 2D models to real systems.

\subsection{Human-made and natural quantum 2D systems}

General understanding of two-dimensional many-body phenomena is important for at least four physical systems:

(i) high-temperature layered cuprate superconductors, where the Cooper pair size is comparable to the interparticle distance, as in the strongly-interacting regime of the Bose-to-Fermi crossover problem. Earlier cited models~\cite{Giorgini2D2011,Fermi2DAbInitioLattice2015,Fermi2DExactGS2015,DiffusionMC2015,Fermi2DEOSandPressureParish2014} account for the \textit{s}-wave coupling only, while the \textit{d}-wave symmetry~\cite{MKaganDPairing1992,MKagan2DtJ1994} dominates in the superconducting phase of the cuprates. Despite that, the models with \textit{s}-wave coupling are also important since the \textit{s}-wave symmetry has been detected in the pseudogap phase of the cuprate superconductors~\cite{sWavePseudogap2012};

(ii) neutron stars, where a number of mixed pasta phases may exist~\cite{PethickNuclPasta1983,VoskresenskyNuckPasta2005,VoskresenskyNuclPasta2006}, one of which is the
nuclear lasagna phase, where the kinematics of particles in predominantly two-dimensional. This phase may be responsible for limiting the rotation period of pulsars~\cite{NuclearPasta2013};

(iii) helium thin films and $^3$He submonolayers~\cite{MKagan2DHe1994};

(iv) ultracold atoms in the 2D geometry.

\subsection{Scope of the review}

This review is devoted to the physics of the Fermi-to-Bose crossover in quantum gases.
Preparation of the ultracold Fermi gases with 2D kinematics is described in section~\ref{sec:UltracoldSpin2D}. Two-body interactions is the subject of section~\ref{sec:Interactions}. This section also discusses the difference between interactions via 2D and 3D potentials as well as mapping one interaction type onto the other. Experimental methods for degeneracy detection and thermometry are reviewed in section~\ref{sec:Thermometry}. Mean-field approaches to the description of the gases in the 3D and 2D Bose-to-Fermi crossover are discussed in section~\ref{sec:MeanField}. Observation of the Fermi-to-Bose crossover is presented in section~\ref{sec:Pressure}. Critical phenomena including Bose condensation, BKT transition, and mean-field description of the latter are discussed in section~\ref{sec:CriticalPhenomena}.
Equation of state (EoS) is the subject of section~\ref{sec:Models}, which also includes simple and advanced models, possible scale invariance specific to 2D gases, and experimental tests of the EoS.
Criteria of two-dimensionality are revisited on the basis of recent experimental progress in section~\ref{sec:Criteria2D}. Spin-imbalanced Fermi gases are the subject of section~\ref{sec:SpinImbal}. We conclude in section~\ref{sec:Concl}.

\section{Ultracold atomic Fermi gases with 2D kinematics}\label{sec:UltracoldSpin2D}
\subsection{Cooling and trapping ultracold atoms}

Experiments with ultracold quantum gases allowed to observe states of matter and effects that previously were just a matter of theoretical discussions:
Bogolubov weakly interacting Bose gas~\cite{WiemanBEC1995};
Bertsch matter~\cite{OHaraScience}, which is a Fermi gas with infinite \textit{s}-wave scattering length and nearly zero interaction range;
crossover between the fermionic Bardeen--Cooper--Schrieffer superfluidity and Bose condensation~\cite{Grimmbeta};
Tonks--Girardeau fermionization of a one-dimensional Bose gas~\cite{BlochTonksGirardeau2004};
and Efimov trimers~\cite{GrimmEfimov2006}.
The ultracold gases are used for quantitative tests of theories that are applicable to condensed matter, high-energy, and nuclear physics~\cite{BlochLowDReview2008,PitaevskiiFermiReview}.

The success of experiments is facilitated by a set of conditions:

-\,in ultracold gases, uncontrolled impurities are absent since the gas preparation is done by spectroscopic methods that are sensitive not only to chemical elements but also to isotopes;

-\,both the interactions and the spin composition may be tuned smoothly and reversibly;

-\,kinematic dimensionality is controlled;

-\,measurements are performed directly: It is possible to instantaneously image the density distribution owing to absorption of light by the atomic gas, to measure thermodynamic properties, to observe the momentum space distribution and the difference between the order-parameter phases of subsystems.

\begin{figure}[htb!]
\begin{center}
\includegraphics[width=\linewidth]{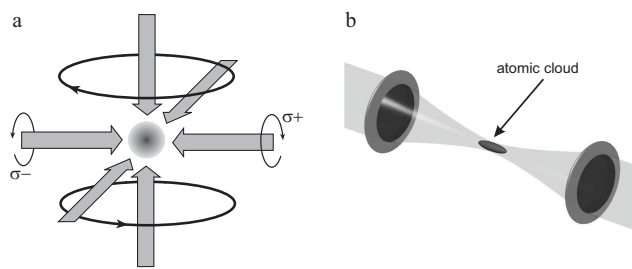}
\end{center}
\caption{(a) Trapping and cooling of an atomic gas in a magneto-optical trap. Six trapping beams, two magnetic coils with opposite currents, and the atomic cloud are shown. (b)~Confinement of an atomic cloud in optical dipole trap formed by a focus of a laser beam.}
\label{fig:Traps}
\end{figure}

Lithium-6 is presently the most popular atom in the experimental studies of the 2D Fermi-to-Bose crossover~\cite{Thomas2D,ValeCloudSize2011,Zwierlein2DPairing2012,FermiBose2DCrossover,Jochim2DPairCondensation2015,SpinImbalanced2D2016} while potassium-40 is another option~\cite{Kohl2DRF}. Achievement of quantum degeneracy requires cooling to temperature $T$ of micro- to nanokelvin range~\cite{OnofrioCoolingReview2016}. In the simplest case, the atomic gas is prepared in two stages, taking few seconds each. At the first stage, trapping and cooling of atoms is performed using laser radiation whose frequency is nearly resonant to an atomic electric-dipole transition~\cite{LetokhovReview2000}. The gas is collected in a magneto-optical trap from either an atomic beam or ambient vapor. The trap is schematically shown in figure~\ref{fig:Traps}a.
By means of the viscous light pressure force, whose value is adjusted by the Zeeman shift, the atomic gas is gathered near the point where the magnetic field is zero.

The magneto-optical trap collects from millions to billions of atoms. Their state is far from quantum degeneracy. For example, for lithium, the phase space density is just $\sim10^{-6}$. Further cooling is impossible due to the resonant light used at this stage. Due to re-emission of photons there is a temperature minimum which is few tens of microkelvin for lithium.
Besides that, further increase of the density is prevented by the light pressure of the atoms upon each other, appearing due to the re-emission.
Therefore, the next stage of cooling is needed. At the start of this stage, the resonant light fields are instantaneously extinguished and the atomic gas finds itself in a conservative potential, provided by an off-resonant optical dipole trap~\cite{OptTrapReview2000}. In the simplest case, such trap is formed in a focus of a laser beam as shown in figure~\ref{fig:Traps}b. The radiation frequency is tuned far below the electrodipole transitions in the atom, which is held in potential $V=-(1/2)\bi{d}\cdot\bi{E}$, where $\bi{E}$ is the electric field, while factor $1/2$ appears since the electric dipole moment $\bi{d}\propto\bi{E}$ is induced by the field. The potential is conservative due to the large detuning between the laser frequency and the atomic transition. The magnetic field is a free parameter and may be used for controlling the interaction between the particles (see section~\ref{sec:Feshbach}).

Further cooling is done by means of evaporation~\cite{LeCooling}. Particles with highest energies leave the trap, while the remaining particles collide and try to form the equilibrium distribution at a lower temperature. This repopulates the higher momentum states, which again brings about losses of the energetic particles. To speed up the evaporation, the trap depth is slowly decreased. As a result, the cooling brings the gas into a degenerate state with phase space density $\sim1$.

The ultracold quantum gases are extremely dilute. The interparticle distance from hundreds of nanometers to microns, which is much larger than the intermolecular distance of 3 nm in the air and the typical scale of the interatomic potentials $r_0\sim0.1$--$1$ nm. At the same time, it is possible to speak about collective behavior typical to fluids: Parts of the system feel each other at large distances due to appearance of a collective wave function or due to the Pauli exclusion principle; similar to the fluids, strong interparticle interactions, whose energy is comparable to the kinetic energy, may appear~\cite{Pitaevskii200letLandau2008eng}. Finally, the weakly-interacting Fermi gas at low temperatures is a Fermi liquid in the Landau sense (see, however nontrivial corrections to the Landau Fermi liquid theory in 2D \cite{MKagan2DFermi1993}). Moreover, this is one of the few systems where the Fermi-liquid parameters are computed from first principles~\cite{LevinFermiLiquid}.

\subsection{Spin states of Fermi atoms}

\begin{figure}[htb!]
\begin{center}
\includegraphics[width=3.0in]{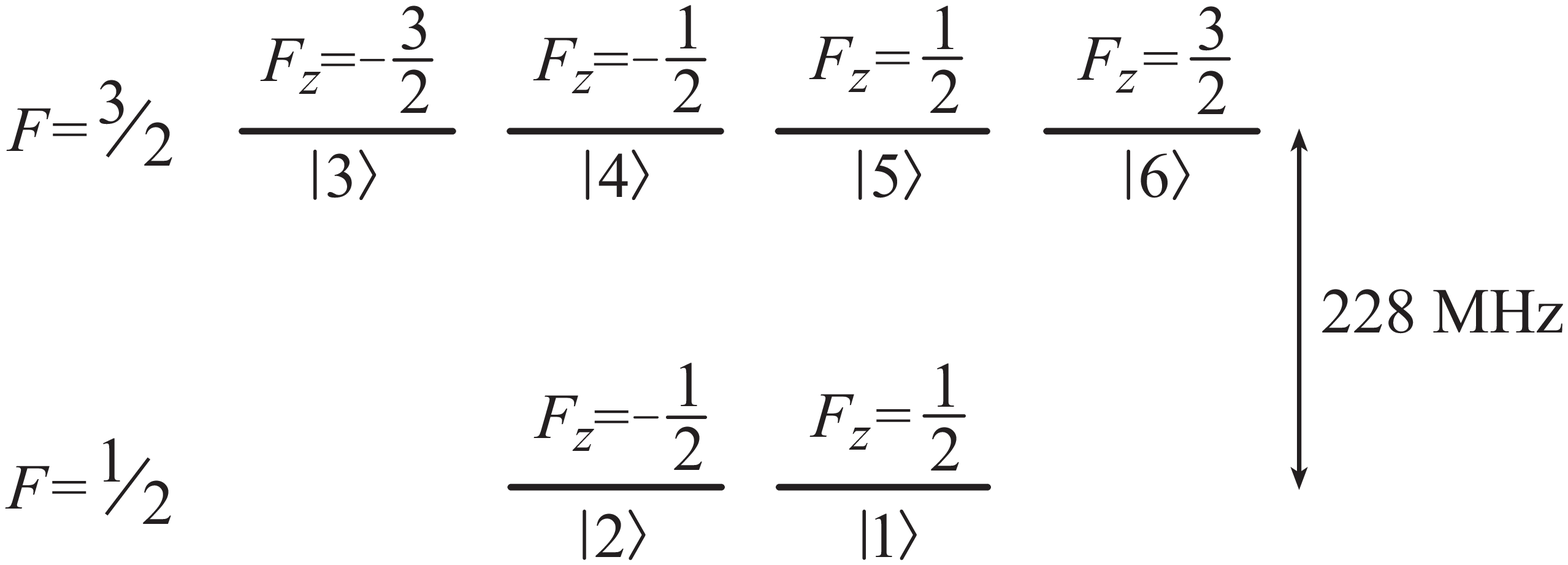}
\end{center}
\caption{States of lithium-6 in zero magnetic field, $B=0$, which correspond to the ground state of the valence electron $2^2$S$_{1/2}$. The states are numbered in the order of increasing energy in magnetic field.}
\label{fig:HyperfineZeroB}
\end{figure}

The spin degrees of freedom are fundamental to the properties of Fermi systems. The role of spin in atomic gas is played by the internal states of the atom, which are discussed here on the example of lithium-6. The states of lithium-6 corresponding to the ground-state orbital of the single valence electron $2s^1$ are shown in figure~\ref{fig:HyperfineZeroB}.
These states differ by the mutual orientation of the valence electron spin $S=1/2$ and the nuclear spin $I=1$. In the absence of the magnetic field, the states may be described in the basis of the total angular momentum operator  $\hat{\bi{F}}=\hat{\bi{S}}+\hat{\bi{I}}$. The balanced mixture of states $|1\rangle$ and $|2\rangle$ is an analog of the gas of spin-up and spin-down electrons in a solid-state system.
Problems requiring more spin diversity may be implemented using a gas of lithium-6, as one may see from figure~\ref{fig:HyperfineZeroB}.

In the experiment, an equal mixture of states $|1\rangle$ and $|2\rangle$ is used. In the external magnetic field $B$, which is turned on for controlling the interactions, the states are expressed in basis $|S_z,I_z\rangle$~\cite{HuletLiCollisions1998}:
\begin{eqnarray}
&&|1\rangle=\cos\theta_+\left|-\frac12,1\right\rangle-\sin\theta_+\left|\frac12,0\right\rangle,\label{eq:GroundState1}\\
&&|2\rangle=\cos\theta_-\left|-\frac12,0\right\rangle-\sin\theta_-\left|\frac12,-1\right\rangle,\label{eq:GroundState2}
\end{eqnarray}
where
\begin{eqnarray}
&&\sin\theta_\pm=\frac1{\sqrt{1+(Z^\pm+R^\pm)^2/2}},\nonumber\\
&&Z^\pm=\frac{2\mu_{\mathrm{B}}B}\alpha\pm1/2,\quad R^\pm=\sqrt{(Z^\pm)^2+2},\nonumber
\end{eqnarray}
$\mu_{\mathrm{B}}$ is the Bohr magneton, and $\alpha/(2\pi\hbar)=152.1$~MHz is the hyperfine interaction constant. In the field $B=600$--$1000$~G typical for the experiments, first terms of superpositional states (\ref{eq:GroundState1}) and (\ref{eq:GroundState2}) are dominant. These terms correspond to projection $S_z=-1/2$. At $B=800$~G, for example, $\sin^2\theta_\pm=0.002$. Despite that, the part of the state corresponding to projection $S_z=1/2$ is principal to the tunability of the interaction.

\subsection{2D kinematics for noninteracting fermions in anisotropic traps}

\begin{figure*}[htb!]
\begin{center}
\includegraphics[width=0.9\linewidth]{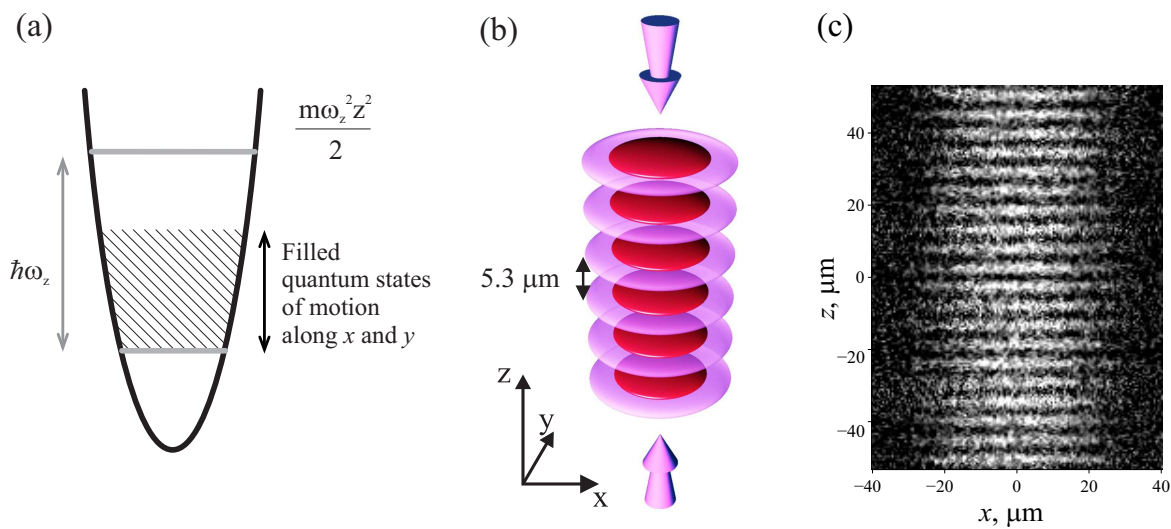}
\end{center}
\caption{(a) Two-dimensional ideal Fermi gas at $T=0$ whose motion is quantized along $z$ and nearly free along $x$ and $y$. (b) Confinement of two-dimensional gas clouds in anti-nodes of a standing electromagnetic wave. The gas is shown in dark red, while the intensity of the radiation forming the trap is shown in light purple. (c) Image of the clouds along the $y$ direction. From \cite{UFN2016eng} \copyright Uspekhi Fizicheskikh Nauk 2016.} \label{fig:TrapAnd2D}
\end{figure*}

Two-dimensional kinematics may be achieved in a strongly-anisotropic parabolic potential
\begin{equation}
V(x,y,z)=\frac{m\omega_z^2z^2}2+\frac{m\omega_x^2x^2}2+\frac{m\omega_y^2y^2}2
\label{eq:OptLattice}
\end{equation}
with $\omega_z\gg\omega_\perp\equiv\sqrt{\mathstrut\omega_x\omega_y}$, $\omega_x\sim\omega_y$, and $m$ being the atom mass. Due to the anisotropy it is possible to place the absolute majority of the atoms into the ground state of motion along $z$, while according to the Pauli exclusion principle, the fermions populate many states of motion in the $xy$ plane as shown in figure~\ref{fig:TrapAnd2D}a.
As a result, the gas is kinematically two-dimensional.

Experimentally a series of such traps may be formed by anti-nodes of an electromagnetic standing wave as schematically shown in figure~\ref{fig:TrapAnd2D}b. The radiation frequency has to be far below that of the electrodipole transitions.
An image of the clouds taken along the plane of motion $xy$ is shown in figure~\ref{fig:TrapAnd2D}c, where each light strip is a separate 2D system.
Here the trapping wavelength is $\lambda=10.6$ $\mu$m, which gives 5.3 $\mu$m separation between the antinodes and, as a result, each cloud is resolved. For imaging, the clouds are shined upon by monochromatic radiation of wavelength 671 nm, which is resonant to the electrodipole transition $2$S$_{1/2}\rightarrow2$P$_{3/2}$ of lithium-6 atom. As a result of the resonant absorption, a shadow appears and is further projected onto a charge coupled device. From the absorption image, the gas density distribution integrated along $y$ is reconstructed~\cite{Fermi2D,FermiBose2DCrossover}. Such photographing is selective to the internal atomic state. The imaging destroys the state of the system due to the energy input, which is much larger than the kinetic energy.
Preparation of few tens of clouds in identical conditions allows for, firstly, averaging the data over the ensemble for the sake of noise suppression and, secondly, observing the interference.

In figure~\ref{fig:TrapAnd2D}c, the gas clouds are in potential with frequencies $\omega_x/(2\pi)=\omega_y/(2\pi)=102$~Hz, $\omega_z/(2\pi)=5570$~Hz. Here factor $1/(2\pi)$ accounts for the conversion from s$^{-1}$ to Hz. The number of atoms in a single cloud per spin state is $N=660$, which gives the Fermi energy $E_{\mathrm{F}}=\hbar\omega_\perp\sqrt{2N}=0.67\hbar\omega_z$ and together with the deep degeneracy nearly excludes thermal population of the excited states of motion along $z$.

For zero-temperature noninteracting fermions, the two-dimensionality condition is $E_{\mathrm{F}}<\hbar\omega_z$, while interactions make such condition less trivial.
Two-body interaction mixes kinematic states of motion along $z$.
From the standpoint of a single atom, therefore, even small two-body interaction brings about some deviation from the 2D kinematics.
In the absence of many-body interactions, however, the center-of-mass kinematics of the pair is 2D due to the nearly harmonic shape of the potential in the $xy$ plane. Conditions of two-dimensionality have been discussed in \cite{Fermi2D,Vale2DCriteria2016,PressureProfile2016}. We shall revisit these conditions in section~\ref{sec:Criteria2D}, which will include discussion of one-, two-, and many-body effects and recent experiments.

\section{Tunable two-body interactions}\label{sec:Interactions}
\subsection{$T$-matrix theory of two-body scattering in 3D and 2D for short-range interactions}
\subsubsection{The 3D vacuum $T$-matrix}

In the 3D case for symmetric attractive potentials there is a threshold for the formation of the bound state when the depth of the potential well $|U|$ exceeds the effective width of the well $W\sim\hbar^2/(mr_0^2)$, where $m$ is the particle mass and $r_0$ is the range of the interaction.
As an example one may take the exponentially decaying potential $U(r)=U\exp(-r'/r_0)$ (here and further primed values are used for coordinates in the center-of-mass reference frame of two interacting particles).
The attractive Hubbard model with contact interaction on the lattice is another example, where $W$ is the bandwidth, $U$ is the Hubbard onsite attraction, and the role of the interaction range ($r_0$) is played by the intersite distance $d$.
For $|U|\ll W$ we are in the weak-coupling case (weak-coupling Born approximation) and there is no bound state of the two particles in real space. The \textit{s}-wave scattering length is negative and given by
\begin{equation}
a_{\mathrm{3D}}=\frac{mU_0}{4\pi\hbar^2}<0
\end{equation}
where $U_0\sim Ur_0^3$ is a zeroth Fourier component of the interaction. In this case only extended \textit{s}-wave Cooper pairing is possible in the BCS domain~\cite{MKaganBook2013}.

If the interaction is strong enough (and the Born parameter $|\gamma|=m|U|r_0^2/(4\pi\hbar^2)$ becomes close to unity) the \textit{s}-wave scattering length and the corresponding 3D vacuum  $T$-matrix for zero total energy and zero total momentum $\bi{K}\equiv\bi{p}_1+\bi{p}_2$ reads:
\begin{equation}
a_{\mathrm{3D}}=\frac{mT(E=0,\bi{K}=0)}{4\pi\hbar^2}\simeq\frac{-|\gamma|r_0}{1-|\gamma|},
\end{equation}
signalling the appearance of a bound state for $|\gamma_{\mathrm{c}}|=1$. Note that if the Born parameter is close to but smaller than one [$(1-|\gamma|)/|\gamma|\ll1$], we are in the resonance situation $|a_{\mathrm{3D}}|\gg1$ and get the shallow virtual bound state $E_{\mathrm{b}}^{\mathrm{3D}}=\hbar^2/(ma_{\mathrm{3D}}^2)$.

For $|\gamma_{\mathrm{c}}|=1$ (at the threshold) an expression for the scattering length $a_{\mathrm{3D}}$ has a pole. For $|\gamma|>1$ we are in the BEC domain of local (compact) pairs or molecules (dimers) and the expression for the \textit{s}-wave scattering length should be modified. For the bound-state problem (for the negative energies $E<0$) we should use the \textit{s}-wave scattering amplitude $f_0(E)$ which reads:
\begin{equation}
|f_0(E)|=\frac{|\gamma|r_0}{1-|\gamma|+\sqrt{m|E|r_0^2}}.
\end{equation}
Above the threshold (but not very far away from it) we have a two-particle bound state with an energy:
\begin{equation}
|E_{\mathrm{b}}|\simeq\frac{\hbar^2}{mr_0^2}\left(\frac{|\gamma|-1}{|\gamma|}\right)^2.
\end{equation}
It is convenient to represent the binding energy as $|E_{\mathrm{b}}^{\mathrm{3D}}|=\hbar^2/(ma_{\mathrm{3D}}^2)$ and thus to introduce repulsive scattering length $a_{\mathrm{3D}}=|\gamma|r_0/(|\gamma|-1)>0$ above the threshold. For $|\gamma_{\mathrm{c}}|=1$ the scattering length diverges. Correspondingly $1/a_{\mathrm{3D}}=0$ and we are in the unitary limit. Slightly above the threshold for $(|\gamma|-1)/|\gamma|\ll1$ we have a shallow bound state $|E_{\mathrm{b}}^{\mathrm{3D}}|=\hbar^2/(ma_{\mathrm{3D}}^2)\ll\hbar^2/(mr_0^2)$ and thus $a_{\mathrm{3D}}\gg r_0$. In the strong-coupling case for $|U|\gg W$ the binding energy $|E_{\mathrm{b}}|\sim|U|$ and we have deep levels with $a_{\mathrm{3D}}\ll r_0$.

\subsubsection{The 2D vacuum $T$-matrix}\label{sec:TMatrix2D}

In the 2D case (in contrast to the 3D situation) the bound state of two particles in vacuum appears already for infinitely small symmetric attractive potential and the threshold for its appearance is absent. Thus even on the BCS-side of the crossover we always have the coexistence of the  two phenomena, the Cooper pairing in momentum space and pairing of two particles in vacuum (in real space). The BCS--BEC crossover is governed by the ratio between the binding energy of the molecule (it is $|E_{\mathrm{b}}|/2$ for one particle in the molecule) and Fermi energy $\varepsilon_{\mathrm{F}}$.
If we again use the convenient notation for the binding energy $|E_{\mathrm{b}}|=\hbar^2/(ma^2)$, then $2\varepsilon_{\mathrm{F}}/|E_{\mathrm{b}}|=a^2k_{\mathrm{F}}^2$ and $\sqrt{2\varepsilon_{\mathrm{F}}/|E_{\mathrm{b}}|}=ak_{\mathrm{F}}$.
Here $\varepsilon_{\mathrm{F}}=\hbar^2k_{\mathrm{F}}^2/(2m)$ is the local Fermi energy and $\hbar k_{\mathrm{F}}$ is the Fermi momentum;
in 2D $\varepsilon_{\mathrm{F}}=2\pi\hbar^2n/m$, where $n$ is the local numerical density per spin component.
The corresponding vacuum $T$-matrix for the negative total energy $E<0$ reads:
\begin{eqnarray}
f_0(E<0)=\frac{mT(E=-|E|,\bi{K}=0)}{4\pi\hbar^2}\nonumber\\
=\frac{-\gamma}{1-\gamma\ln\left(\frac W{|E|}+1\right)},
\label{eq:TMatrix2D}
\end{eqnarray}
where $\gamma=m|U_0|/(4\pi\hbar^2)$ is the effective Born radius and $W=\hbar^2/(mr_0^2)$ is the effective width of the potential well. In (\ref{eq:TMatrix2D}), $U_0\sim Ur_0^2$ is the zeroth Fourier component of the potential in 2D. As a result the pole of the $T$-matrix which corresponds to the energy of the bound state in (\ref{eq:TMatrix2D}) reads:
\begin{equation}
|E_{\mathrm{b}}|=\frac{\hbar^2}{ma^2}=\frac W{\exp\left(\frac1\gamma\right)-1}.
\end{equation}
Correspondingly it is useful to express the scattering amplitude via the binding energy $|E_{\mathrm{b}}|$ which yields the following elegant expression:
\begin{equation}
f_0(E<0)= \frac1{\ln\left(\frac{W+|E|}{W+|E_{\mathrm{b}}|}\frac{|E_{\mathrm{b}}|}{|E|}\right)}.
\label{eq:f0viaEbFor2D}
\end{equation}
(\ref{eq:f0viaEbFor2D}) correctly describes both the situation with shallow levels for which $|E|\sim|E_{\mathrm{b}}|\ll W$ and with deep levels where $|E|\sim|E_{\mathrm{b}}|\gg W$.

For deep levels
\begin{equation}
f_0(E<0)\simeq\frac{|E|\,|E_{\mathrm{b}}|}{W(|E_{\mathrm{b}}|-|E|)}.
\end{equation}
Since $\gamma\gg1$ for deep levels, we can expand $\exp(1/\gamma)\simeq1+1/\gamma$. Thus
\begin{equation}
|E_{\mathrm{b}}|\simeq\gamma W=\frac{|U_0|}{4\pi r_0^2}\sim U.
\label{eq:DeepEb2D}
\end{equation}

For shallow levels we get:
\begin{equation}
f_0(E<0)\simeq\frac1{\ln(E_{\mathrm{b}}/E)}.\label{eq:f02DShallow}
\end{equation}
In the shallow-level case $\exp(1/\gamma)\gg1$ (or equivalently $\gamma\ll1$) and hence the binding energy reads:
\begin{equation}
|E_{\mathrm{b}}|=W\exp\left(-\frac1\gamma\right).
\end{equation}
A more conventional scattering amplitude $f_{\mathrm{2D}}$ may be obtained from (\ref{eq:f02DShallow}) by analytic continuation to positive energies $E=\hbar^2q^2/m$.
In (\ref{eq:f02DShallow}), the binding energy may be expressed via the 2D scattering length $a_2$ ($a_2\rme^{\gamma_{\mathrm{E}}}/2\equiv a$); $E_{\mathrm{b}}=-4\hbar^2/(ma_2^2e^{2\gamma_{\mathrm{E}}})$, where $\gamma_{\mathrm{E}}\simeq0.577$ is Euler's constant. As a result one obtains:
\begin{equation}
f_{\mathrm{2D}}(q,a_2)=4\pi f_0(E>0)=-\frac{2\pi}{\ln[qa_2\rme^{\gamma_{\mathrm{E}}}/(2\rmi)]},
\label{eq:f2D}
\end{equation}
where $\hbar q\equiv|\bi{p}_1-\bi{p}_2|/2$ is the relative momentum of two colliding atoms, while $\bi{p}_1$ and $\bi{p}_2$ are their momenta in the laboratory reference frame. Note that $f_{\mathrm{2D}}$ is different from the standard 2D scattering amplitude~\cite{LandauQuantumEng}, which equals $-f_{\mathrm{2D}}/\sqrt{8\pi q}$.
The scattering wavefunction is \cite{LandauQuantumEng}
\begin{equation}
\psi_\perp(\boldsymbol{\rho'})\simeq\exp(\rmi\bi{q}\cdot\boldsymbol{\rho'})- f_{\mathrm{2D}}\frac{\exp(\rmi q\rho'-\rmi\pi/4)}{\sqrt{\mathstrut8\pi q\rho'}},\label{eq:ScattWavefunc}
\end{equation}
where $\boldsymbol{\rho'}$ is the vector between the two atoms in the $xy$ plane. The \textit{s}-wave part of the wavefunction of two unbound atoms at large distances asymptotically behaves as $\psi_\perp(\rho')\propto\ln\rho'/a_2$, where $\rho'\equiv|\boldsymbol{\rho'}|$.

\subsection{2-body scattering for 2D kinematics and 3D contact interaction}

The situation with atomic gases in the tight parabolic trap bears similarities and differences with the case where both kinematics and the interaction potential are purely 2D. In the experiments, the range $r_0$ of the 3D potential is much smaller than the quantization size $l_z\equiv\sqrt{\hbar/2m\omega_z}$. The interaction potential may be regarded as the three-dimensional $\delta$-function on the scale of the problem. As a result, the interactions are quasi-two-dimensional rather than two-dimensional because at distances $\ll l_z$ the wavefunction of colliding atoms is determined by three-dimensional scattering length $a_{\mathrm{3D}}$, which is the major difference with the purely 2D scattering problem.

The similarity appears in the form of the scattering wavefunction, which at large distances is the same as (\ref{eq:ScattWavefunc}), except the scattering amplitude $f_{\mathrm{2D}}$ is to be replaced by the expression which is specific to the quasi-2D scattering~\cite{Shlyapnikov2DScattering2001}
\begin{equation}
f=f_{\mathrm{Q2D}}(q,a_{\mathrm{3D}},l_z)\equiv\frac{2\pi}{\sqrt\pi l_z/a_{\mathrm{3D}}+w(q^2l_z^2)/2},
\label{eq:f2ho}
\end{equation}
where the function $w(\xi)$ is defined via the limit
\begin{eqnarray}
&&w(\xi)\!\equiv\!\lim_{J\rightarrow\infty}\!\left[\sqrt{\frac{4J}\pi}\ln\frac J{e^2}-\right.\nonumber\\ &&\left.\sum_{j=0}^J\frac{(2j-1)!!}{(2j)!!}\ln(j-\xi-i0)\right]\!\!.
\label{eq:wOfxi}
\end{eqnarray}

\subsection{Parametrization of the quasi-2D scattering via the 2D scattering length}\label{sec:Scattering2D}

The theory of two-dimensional many-body systems traditionally uses the two-dimensional \textit{s}-wave scattering length $a_2$ \cite{Bloom1975,FermiLiquid2D1992}. Parametrization of quasi-2D scattering in terms of $a_2$ is important for relating ultracold-atom experiments to the body of theoretical models formulated for purely 2D systems.

Quasi-2D interaction is parametrized via $a_{\mathrm{3D}}$, $l_z$, and $q$ since $f_{\mathrm{Q2D}}$ is a function of those variables. The wavefunction structure (\ref{eq:ScattWavefunc}), which is common for both pure 2D and quasi-2D collisions, opens the way to parametrize quasi-2D collisions using $a_2$. Since the only difference between 2D and quasi-2D is which scattering amplitude ($f_{\mathrm{2D}}$ or $f_{\mathrm{Q2D}}$) to substitute into (\ref{eq:ScattWavefunc}), the solution of the equation
\begin{equation}
f_{\mathrm{Q2D}}(q,a_{\mathrm{3D}},l_z)=f_{\mathrm{2D}}(q,a_2)
\label{eq:Finda2}
\end{equation}
gives the $a_2$ value for parametrizing the quasi-2D collision~\cite{FermiBose2DCrossover}.

In the limit $q\rightarrow0$ this approach yields known expression~\cite{Shlyapnikov2DScattering2001}:
\begin{equation}
a_2\simeq2.96\,l_z\rme^{-l_z\sqrt\pi/a_{\mathrm{3D}}},\label{eq:a2atZeroEnergy}
\end{equation}
where, as one may note, no discontinuity appears at resonance ($a_2=2.96l_z$ as $a_{\mathrm{3D}}$ jumps from $-\infty$ to $+\infty$). In a many-body problem, the momentum $\hbar q$ differs from 0 and may be estimated as $\hbar q=\sqrt{\mathstrut2\tilde\mu m}$ from the chemical potential $\tilde\mu$ that does not include the two-body binding energy~\cite{FermiBose2DCrossover}. This estimate is exact for deeply degenerate weakly interacting Fermi gases where the colliding particles are on the Fermi surface ($q=k_{\mathrm{F}}$).

In this section and further, $a_2$ is related to the 3D scattering parameters in the limit $r_0\rightarrow0$. The role of a finite interaction range is discussed in \cite{Parish2DFermiReview2015}.

There is an alternative way of finding the corresponding value of the 2D scattering length and binding energy, which are respectively denoted $\tilde a_2$ and $\tilde E_{\mathrm{b}}$ for this case. The calculation is based on equation~\cite{BlochLowDReview2008}:
\begin{equation}
\frac{l_z}{a}=\int_0^\infty\frac{du}{\sqrt{8\pi u^3}}\left(1- \frac{e^{-u|\tilde E_{\mathrm{b}}|/\hbar\omega_z}}{\sqrt{(1-e^{-2u})/2u}}\right),
\label{eq:Eb}
\end{equation}
from where $\tilde E_{\mathrm{b}}$ is found. Subsequently $\tilde a_2$ is obtained from
$\tilde E_{\mathrm{b}}=-4\hbar^2/(m\rme^{2\gamma_E}\tilde a_2^2)$.

For low binding energies $|\tilde E_{\mathrm{b}}|\ll\hbar\omega_z$ (small negative $a_{\mathrm{3D}}$), the two approaches give nearly identical results ($a_2\simeq\tilde a_2$) in the $q\rightarrow0$ limit. The controversy comes at stronger binding $|\tilde E_{\mathrm{b}}|\gg\hbar\omega_z$ (small $a_{\mathrm{3D}}>0$) \cite{FermiBose2DCrossover}, which is seen by considering the mean field of a uniform 2D gas of atoms. The leading-order term is $-2\pi\hbar^2n/[m\ln(a_2\sqrt n)]$~\cite{HardCore2DBosons1971}. Plugging in $a_2$ derived from (\ref{eq:Finda2}), one obtains the mean-field value $\sqrt{4\pi}\hbar^2na_{\mathrm{3D}}/(ml_z)$ in agreement with \cite{Pricoupenko2DMFvsVariational2004},
while (\ref{eq:Eb}) yields much larger $\tilde a_2\gg a_2$ overestimating the mean field.

\subsection{Tuning the interactions by means of a magnetic Fano--Feshbach resonance}\label{sec:Feshbach}

Formula~(\ref{eq:a2atZeroEnergy}) shows that the effective 2D scattering length $a_2$ may be controlled by changing either $a_{\mathrm{3D}}$ or $\l_z$. The 3D scattering length $a_{\mathrm{3D}}$ may be tuned using magnetically controlled Fano--Feshbach resonances~\cite{FeshbachReview2010}.

\begin{figure}[h!]
\begin{center}
\includegraphics[width=2.5in]{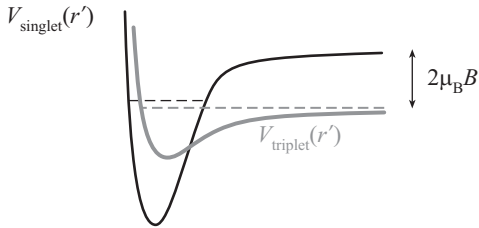}
\end{center}
\caption{The Fano--Feshbach resonance: unbound pair of particles in the triplet channel with the energy marked by the gray dashed line comes into resonance with the energy of the singlet channel bound state (black dashed line). Relative energies of the bound and unbound states in the figure correspond to $a_{\mathrm{3D}}<0$.}
\label{fig:FeshbachPrincipal}
\end{figure}

Here the fundamentals of the Fano--Feshbach resonance are reviewed on the example of $^6$Li atoms. Since the focus is on controlling $a_{\mathrm{3D}}$, the 3D kinematics is discussed.
In the \textit{s}-wave approximation, the two colliding atoms have to be in orthogonal internal states, \textit{e.~g.}, $|1\rangle$ and $|2\rangle$ for $^6$Li.
The appearance of the resonance and tunability of the scattering length is illustrated in figure~\ref{fig:FeshbachPrincipal}.
At the interaction of two univalent atoms, the valence electrons are in a superposition of the triplet state with collinear spins and singlet state with opposite spins.
Different potentials of interparticle interaction, $V_{\mathrm{triplet}}(r')$ and $V_{\mathrm{singlet}}(r')$,  respectively correspond to the triplet and singlet state of the electrons. These potentials are shown in figure~\ref{fig:FeshbachPrincipal}.
The external magnetic field $B$ shifts the zero level of the triplet-state kinetic energy since the state has large magnetic moment, about $\simeq2\mu_{\mathrm{B}}$. If the energy of the unbound state of the pair, shown by the gray dashed line, becomes equal to the singlet-channel bound-state energy (black dashed line) than a scattering resonance appears: In the case of zero kinetic energy, the scattering length diverges to $\infty$. By detuning bound and free state from each other, any desired scattering length between $-\infty$ and $\infty$ may be obtained. Near the resonance, an approximate formula may be used to the \textit{s}-wave scattering length of two unbound particles:
\begin{equation}
a_{\mathrm{3D}}(B)=a_{\mathrm{bg}}\left(1+\frac\Delta{B-B_0}\right),
\end{equation}
where $B_0$ and $\Delta$ are the center location and the width of the resonance respectively, while $a_{\mathrm{bg}}$ is the background scattering length stemming from the triplet channel alone.
For the majority of experiments discussed below, the Feshbach resonance with parameters $B_0=832$~G, $\Delta=262$~G, $a_{\mathrm{bg}}=-1580a_0$ ($a_0$ is the Bohr radius) \cite{JochimNewLiFeshbach2013} is employed. Note that at such big magnetic field, the triplet-state part of the electronic spin dominates in the state of the pair of atoms $|1\rangle$ and $|2\rangle$ [see formulas (\ref{eq:GroundState1}) and (\ref{eq:GroundState2})] since coefficients $\sin\theta_\pm$ are small. The presence of a singlet part, at least small, in the state of electronic spins is principal to the coupling of the channels $V_{\mathrm{singlet}}(r')$ and $V_{\mathrm{triplet}}(r')$, tunability of the interactions, and appearance of the resonance.

In a many-body system, the Fermi atoms are joined into diatomic molecular bosons by tuning the interactions.
For this purpose, the magnetic field smoothly changes from larger values, where the bounds state is above the energy of the free state in the triplet channel (as in figure~\ref{fig:FeshbachPrincipal}), to smaller values. On the bosonic side of the resonance ($B<B_0$), at sufficiently large detuning from the resonance, a Bose condensate of molecules appears.

\section{Degeneracy detection and thermometry}\label{sec:Thermometry}

\begin{figure}[htb!]
\begin{center}
\includegraphics[width=\linewidth]{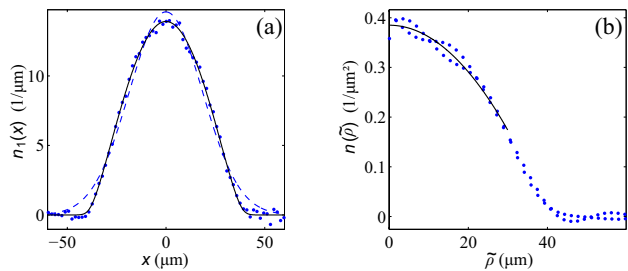}
\end{center}
\caption{(In color online.) (a) Linear numerical density profile $n_1(x)$. Dots: data for $a_2\sqrt{n}=55$, $B=1400$~G, $\omega_x/(2\pi)=94$~Hz, $\omega_y/(2\pi)=141$~Hz, $\omega_z/(2\pi)=6020$~Hz,  and $N=660$. Fits of the Thomas--Fermi~(\ref{eq:n1ThomasFermi}) and Gaussian profile to the data are respectively shown by the solid and dashed curves. (b)~Two-dimensional numerical density distribution in the $xy$ plane, $n(\tilde{\rho})$, found from $n_1(x)$. Dots are the data. The curve is the fit of the parabola $n(\tilde{\rho})=n-\tilde{\rho}^2n''/2$ for finding the central numerical density $n\equiv n(\tilde{\rho}=0)$.}
\label{fig:DensityProfiles}
\end{figure}

The form of the trapped gas density profile serves as a source of information about the temperature.
In figure~\ref{fig:DensityProfiles}a one may see an example of the linear-density profile $n_1(x)=\int n(x,y)\rmd y$.
Absorption images similar to that of figure~\ref{fig:TrapAnd2D}c effectively provide the integration along the line of sight $y$. The $n_1(x)$ data are obtained by integrating such image along $z$ which, for this particular case, resulted in averaging over 30 nearly identical clouds~\cite{FermiBose2DCrossover}.
At high temperature, $T>E_{\mathrm{F}}$, the gas is close to classical, and the density profile is close to the Gaussian distribution. At $T=0$ the edges of the distribution $n_1(x)$ are sharper and, in case of a noninteracting Fermi gas, the numerical density distribution has the form
\begin{equation}
n_1(x)=\cases{\frac{8N}{3\pi R_x}\left(1-\frac{x^2}{R_x^2}\right)^{3/2}&for $x<R_x$,\\
0&for $x>R_x$,}
\label{eq:n1ThomasFermiZeroT}
\end{equation}
where $R_x\equiv\sqrt{2E_{\mathrm{F}}/m\omega_x^2}$ is the Thomas--Fermi radius. At an arbitrary temperature $T$ the numerical density profile of a nearly ideal Fermi gas is expressed as
\begin{equation}
n_1(x)=-\sqrt{\frac{m\omega_\perp}{2\pi\hbar}}\left(\frac T{\hbar\omega_\perp}\right)^{3/2}\mathrm{Li}_{3/2}\left(-e^{\frac\mu T-\frac{m\omega_\perp^2x^2}{2T}}\right),
\label{eq:n1ThomasFermi}
\end{equation}
where $\mathrm{Li}_{3/2}$ is the polylogarithm function of the order 3/2 and the chemical potential is to be found self-consistently, from the constraint $N=\int n_1(x)\rmd x$. By fitting this profile to the data of figure~\ref{fig:DensityProfiles}a one may find the temperature $T$. In experiment the temperatures $T\lesssim0.1E_{\mathrm{F}}$ are accessible, which is $\simeq10$~nK in absolute units.

For comparison, a trial fit by a gaussian curve is shown in figure~\ref{fig:DensityProfiles}a. One may see that the Gaussian is off both at the edges and center. For interacting Fermi and Bose gases, the numerical density profiles differ from profile~(\ref{eq:n1ThomasFermi}). Despite that, since at $T=0$ the dependence of the chemical potential on the density is nearly linear ($\tilde\mu\propto n$) \cite{Kohl2DBreathingMode2012}, the closeness of the density profile to (\ref{eq:n1ThomasFermiZeroT}) firmly indicates a deep degeneracy and a smallness of the temperature with respect to $E_{\mathrm{F}}$ and chemical potential $\tilde\mu=\mu+|E_{\mathrm{b}}|/2$.

\begin{figure}[htb!]
\begin{center}
\includegraphics[width=0.8\linewidth]{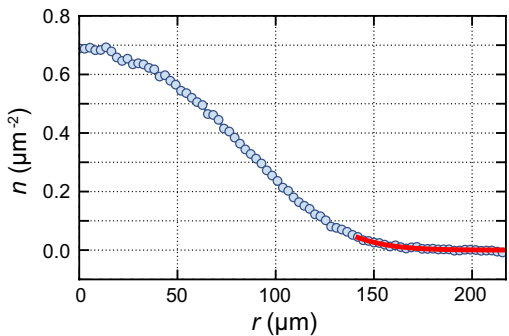}
\end{center}
\caption{(In color online.)  Fit of the in-trap density with the third-order virial expansion. Blue points are the average of 20 experimental images. Red solid line is the virial expansion fit. From \cite{Vale2DThermodynamics2015}.}
\label{fig:ValeVirialThermometry}
\end{figure}

Quantitative thermometry of interacting gases is provided by fitting the density distribution at the cloud edge where the gas is nearly classical~\cite{Vale2DThermodynamics2015,Jochim2DThermodynamics2015} as shown in figure~\ref{fig:ValeVirialThermometry}.
The interactions are taken into account in various orders by means of the virial expansion of the phase space density $n\lambda_{\mathrm{dB}}^2$, where $\lambda_{\mathrm{dB}}=\sqrt{2\pi\hbar^2/(mT)}$ is the thermal de Broglie wavelength:
\begin{equation}
n\lambda_{\mathrm{dB}}^2=\ln(1+e^{\mu/T})+2b_2e^{2\mu/T}+3b_3e^{3\mu/T}+...,
\end{equation}
Here $n(\boldsymbol\rho)$ is the 2D numerical density profile, $\mu(\boldsymbol\rho)=\mu_0-V(\boldsymbol\rho)$ is the local chemical potential including the binding energy, while $\mu_0$ is the trap-center value. The first term of the the right-hand side corresponds to the noninteracting gas, while the second and third term respectively account for the two- and three-body interactions between the fermions.
The expansion parameter is the fermionic fugacity $e^{\mu/T}$, which has to be $\ll1$ in order to keep the expansion valid. The 2nd virial coefficient may be found within the Beth--Uhlenbeck approach~\cite{VirialExpansion1937,Hofmann2014,Vale2DThermodynamics2015,Jochim2DThermodynamics2015}:
\begin{equation}
b_2=e^{|E_{\mathrm{b}}|/T}- \int_{-\infty}^\infty\frac{\exp[-e^s/(2\pi)]\,\rmd s}{\pi^2+[s-\ln(2\pi|E_{\mathrm{b}}|/T)]^2},
\label{eq:VirialExpansion}
\end{equation}
while the 3rd virial coefficient $b_3$ has been calculated in \cite{DrummondVirial2D2010,Parish2DPairs2013}. The fit parameters are $T$ and $\mu_0$.

The virial expansion up to the 3rd order has been used for thermometry in \cite{Vale2DThermodynamics2015}. The gas in this experiment was in the fermionic and strongly-interacting regimes; in all cases the magnetic fields corresponded to the values on the Fermi side of the respective 3D Feshbach resonance. Figure~\ref{fig:ValeVirialThermometry} shows the fit of the data by the third-order virial expansion.

In \cite{Jochim2DThermodynamics2015}, the virial expansion is used up to the 2nd order. Importantly, it is shown that the expansion (\ref{eq:VirialExpansion}) up to the 2nd order is applicable in the whole crossover. In the Bose regime expression (\ref{eq:VirialExpansion}) converges to the ideal-gas Boltzmann distribution since the dimer-dimer interactions are not accounted for. For fugacities that are lesser but comparable to 1, the second-order virial thermometry estimates the temperature from below~\cite{Jochim2DThermodynamics2015}.

In the deeply degenerate regime, the virial-expansion methods may contain errors related to the shrinking of the cloud edge that is used for the fit as well as due to the breakdown of the local density approximation needed for calculating the virial coefficients.

\section{Mean-field approaches to the Fermi-to-Bose crossover}\label{sec:MeanField}

\subsection{Fermi-to-Bose crossover in 3D}
In dilute 3D Fermi gas (with substantially weak attraction between particles $|a_{\mathrm{3D}}|k_{\mathrm{F}}\ll1$) the BCS critical temperature is given by the famous Gor'kov--Melik-Barkhudarov formula~\cite{GorkovCorrection1961}:
\begin{equation}
T_{\mathrm{c}}^{\mathrm{BCS}}=0.28\varepsilon_{\mathrm{F}}\exp\left(-\frac1{|f_0|}\right),
\end{equation}
where a 3D gas parameter $|f_0|=2|a_{\mathrm{3D}}|k_{\mathrm{F}}/\pi$. Correspondingly the BEC critical temperature yields:
\begin{equation}
T_{\mathrm{c}}^{\mathrm{BEC}}\simeq 0.2\varepsilon_{\mathrm{F}}\left(1+1.3a_{2-2}^{\mathrm{3D}}n_{\mathrm{3D}}^{1/3}\right),
\end{equation}
where $n_{\mathrm{3D}}$ is the numerical density of the dimers and the nontrivial corrections to the Einstein formula~\cite{EinsteinCondensation1925} according to Prokof'ev, Svistunov and colleagues are governed by the dimer-dimer scattering amplitude \cite{SvistunovTcShift2001,SvistunovBKT2001}. In exact calculations of Petrov \textit{et al.} \cite{Petrov} and Brodsky \textit{et al.} \cite{MKagan06a2005,MKagan06a2006} the dimer-dimer 3D scattering length is $a_{2-2}^{\mathrm{3D}}=0.6a_{\mathrm{3D}}$.

\begin{figure}[htb!]
\begin{center}
\includegraphics[width=0.9\linewidth]{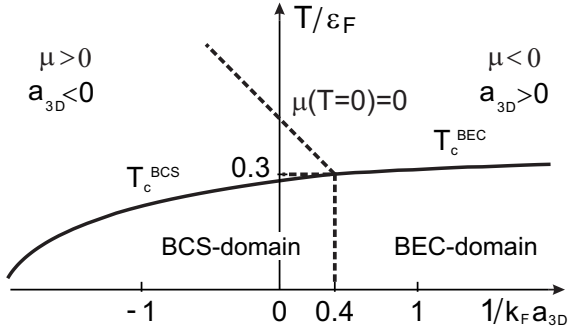}
\end{center}
\caption{Phase diagram of the BCS-BEC crossover in the resonance Fermi gas in 3D (numerical calculations for $T_{\mathrm{c}}$ and $\mu$ versus $1/(k_{\mathrm{F}}a)$ for the first iteration to the selfconsistent $T$-matrix approximation) according to \cite{MKaganCrossover2006,MKaganBook2013}.}
\label{fig:PhaseDiagram}
\end{figure}

The typical phase diagram of the BCS--BEC crossover in 3D resonance Fermi gas is presented on figure~\ref{fig:PhaseDiagram}.
The BCS domain corresponds to the \textit{s}-wave scattering length $a_{\mathrm{3D}}<0$ and the chemical potential $\mu>0$. In the BEC domain vice versa $a_{\mathrm{3D}}>0$, $\mu<0$. In figure~\ref{fig:PhaseDiagram} we plot dimensionless critical temperature as a function of the inverse gas parameter $1/(k_{\mathrm{F}}a_{\mathrm{3D}})$.

\subsection{Fermi-to-Bose crossover in 2D}

In the 2D case on the mean-field level the BCS--BEC crossover was firstly addressed by Miyake~\cite{Miyake1983} (see also important papers~\cite{Randeria2DCrossover1989prl,SchmittRink2DFermiPairing1989}) on the basis of the self-consistent Leggett's theory~\cite{LeggettColloq1980,Leggett}. It is important to mention here also  Fisher--Hohenberg theory~\cite{Hohenberg2DBose1988} for weakly interacting repulsive 2D Bose gas which used the Hartree--Fock ansatz for the chemical potential. According to Miyake the mean-field BCS critical temperature for the BCS domain of extended Cooper pairs is given by:
\begin{equation}
T_{\mathrm{c}}^{\mathrm{BCS}}=\sqrt{2\varepsilon_{\mathrm{F}}|E_{\mathrm{b}}|}.
\end{equation}
Correspondingly for low temperatures $T\ll|E_{\mathrm{b}}|\le2\varepsilon_{\mathrm{F}}$ the chemical potential
\begin{equation}
\mu\simeq\varepsilon_{\mathrm{F}}-|E_{\mathrm{b}}|/2.
\label{eq:muMiyake}
\end{equation}

The mean-field BEC critical temperature according to Fisher and Hohenberg~\cite{Hohenberg2DBose1988} is given by:
\begin{equation}
T_{\mathrm{c}}^{\mathrm{BEC}}=\frac{\varepsilon_{\mathrm{F}}}{4\ln(1/f_{2-2})},
\end{equation}
where according to Petrov, Baranov, and Shlyapnikov \cite{Shlyapnikov2DFermi2003} the dimer-dimer scattering amplitude in 2D case reads:
\begin{equation}
f_{2-2}\sim\frac1{\ln(1.6|E_{\mathrm{b}}|/\varepsilon_{\mathrm{F}})}.
\end{equation}

\begin{figure}[htb!]
\begin{center}
\includegraphics[width=0.9\linewidth]{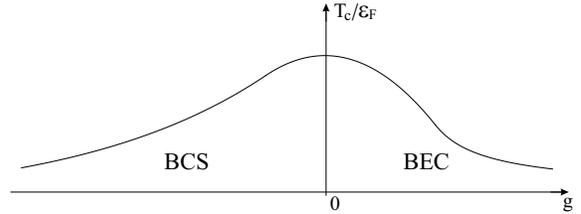}
\end{center}
\caption{Qualitative dependence of the dimensionless critical temperature $T_{\mathrm{c}}/\varepsilon_{\mathrm{F}}$ on the convenient parameter $g=(1/2)\ln[|E_{\mathrm{b}}|/(2\varepsilon_{\mathrm{F}})]$ for the mean-field treatment of the 2D BCS-BEC crossover. The left domain under the curve for the dimensionless critical temperature $T_{\mathrm{c}}/\varepsilon_{\mathrm{F}}$ corresponds to the BCS phase (for this phase $|E_{\mathrm{b}}|<2\varepsilon_{\mathrm{F}}$ and $g<0$), while the right one corresponds to the BEC phase ($|E_{\mathrm{b}}|>2\varepsilon_{\mathrm{F}}$ and $g>0$). The intermediate coupling case with strong fluctuations corresponds to $|E_{\mathrm{b}}|\sim2\varepsilon_{\mathrm{F}}$ and small absolute values of $g$.} \label{fig:PhaseDiagram2D}
\end{figure}

In the Fermi-liquid expansions for normal state and for the BCS--BEC crossover in superfluid state we should use the 2D gas parameter which is usually defined as the \textit{s}-wave scattering amplitude for the absolute value of the total energy $|E|$ equal to $2\varepsilon_{\mathrm{F}}$. Thus we get $f_0(|E|=2\varepsilon_{\mathrm{F}})$ or correspondingly
\begin{equation}
f_0(k_{\mathrm{F}}a)=-\frac1{2\ln(k_{\mathrm{F}}a)}.
\label{eq:GasParameter2D}
\end{equation}
This is a standard definition of the 2D gas parameter both in Fermi gas (with strong repulsion) and Fermi gas with (weak and strong) attraction. If we introduce a convenient variable
\begin{equation}
g\equiv\frac1{2f_0(p_{\mathrm{F}}a)}=\frac12\ln\left(\frac{|E_{\mathrm{b}}|}{2\varepsilon_F}\right)
\label{eq:g2D}
\end{equation}
and plot the dimensionless critical temperature $T_{\mathrm{c}}/\varepsilon_{\mathrm{F}}$ as a function of $g$, then $T_{\mathrm{c}}^{\mathrm{BEC}}/\varepsilon_{\mathrm{F}}\sim1/[4\ln(6.4g)]$, $T_{\mathrm{c}}^{\mathrm{BCS}}/\varepsilon_{\mathrm{F}}\sim2\exp(g)$ and we reproduce the same qualitative picture of the BCS--BEC crossover in figure~\ref{fig:PhaseDiagram2D} as the one presented in figure~\ref{fig:PhaseDiagram}.
Note that at low temperatures $T\ll\varepsilon_{\mathrm{F}}\le|E_{\mathrm{b}}|/2$ the one-particle chemical potential in BEC domain at first (in the intermediate situation $|E_{\mathrm{b}}|\ge2\varepsilon_{\mathrm{F}}$ and for shallow bound states) has the same form as in (\ref{eq:muMiyake}) and contains the factor $\varepsilon_{\mathrm{F}}$. However for deep levels in 2D (when $\varepsilon_{\mathrm{F}}\ll\hbar^2/(mr_0^2)\ll|E_{\mathrm{b}}|\sim|U|$) the chemical potential (neglecting corrections of the order of $\varepsilon_{\mathrm{F}}/W$, $W/|E_{\mathrm{b}}|$) is just $\mu\simeq-|E_{\mathrm{b}}|/2$. This result does not take into account weak dimer-dimer scattering in 2D Fermi gas. When we take this repulsive dimer-dimer interaction into account we finally get:
\begin{equation}
\mu\simeq-\frac{|E_{\mathrm{b}}|}2+\frac{\pi n}{2m}f_{2-2},
\end{equation}
where $n$ is the 2D numerical density of the bosonic dimers. Correspondingly bosonic chemical potential
\begin{equation}
\mu_{\mathrm{Bose}}=2\mu+|E_{\mathrm{b}}|\simeq\frac{\pi n}mf_{2-2}
\end{equation}
as it should be.

Note also that in the BEC regime $|E_{\mathrm{b}}|>2\varepsilon_{\mathrm{F}}$ there is one more characteristic temperature governed by the Saha formula for the thermodynamic equilibrium~\cite{LandauStatFizIeng} in the process $A+B\leftrightarrow AB$. This crossover temperature is given by $T_*=|E_{\mathrm{b}}|/\ln[E_{\mathrm{b}}/(2\varepsilon_{\mathrm{F}})]$. For the temperatures $T_{\mathrm{c}}^{\mathrm{BEC}}<T<T_*$ we are in the strong pseudogap regime corresponding to an interesting phase of a normal (non-superfluid) dilute gas of composed bosons~\cite{MKaganSpectralFunc1998,MKaganSpectralFunc2000,MKaganCrossover2006,StrinatiPseudogap2015}. The pseudogap phenomenon is also pronounced in the intermediate coupling case $|E_{\mathrm{b}}|\sim2\varepsilon_{\mathrm{F}}$ \cite{StrinatiPseudogap2015}.

\section{Observing the Fermi-to-Bose crossover in a deeply degenerate gas by measuring the pressure}\label{sec:Pressure}

The pressure value directly indicates the transition between the Bose and Fermi regimes. Whenever the total pressure of the two spin components $P$ is close to that of an ideal Fermi gas, $P_{\mathrm{ideal}}$, the system is fermionic, while $P/P_{\mathrm{ideal}}\ll1$ indicates the bosonic regime. The ratio $P/P_{\mathrm{ideal}}$ has been measured in \cite{FermiBose2DCrossover} at different interaction values at deeply degeneracy. The preparation of the ultracold gas of $^6$Li in 2 equally populated spin states and tuning the interactions is discussed in secions~\ref{sec:UltracoldSpin2D} and \ref{sec:Interactions} respectively. The schematic of holding a series of 2D clouds is shown if figure~\ref{fig:TrapAnd2D}b, and the images of the trapped clouds are in figure~\ref{fig:TrapAnd2D}c.

For observing the Fermi-to-Bose crossover, the system properties at the center are especially interesting since the gas is mostly degenerate there. Besides that, all known models are constructed for the uniform systems. Therefore, for quantitative comparison, one needs a measurement at the mostly uniform cloud part, which the cloud center is.
While the image of the trapped gas (figure~\ref{fig:TrapAnd2D}c) is taken along the planes of motion and the centers are not seen directly, it is still possible to obtain the pressure and the ratio $P/P_{\mathrm{ideal}}$ for the cloud centers.

The measurement is based on analyzing the linear density profiles $n_1(x)$, such as the one in figure~\ref{fig:DensityProfiles}a, and the use of force balance equation
\begin{equation}
\nabla_\perp P(x,y)=-2n(x,y)\nabla_\perp V(x,y,0),\label{eq:PressureBalance}
\end{equation}
where $V(x,y,0)$ is the trapping potential and $n(x,y)$ is the numerical density per spin state or density of the bosons. Integrating (\ref{eq:PressureBalance}), one finds that for a harmonic potential the central pressure is independent of the interaction: $P=m\omega_\perp N/\pi$, where $N$ is the atom number per spin state. Further, the pressure is normalized to the local Fermi pressure, \textit{i.~e.}, to the ideal Fermi gas pressure at $T=0$ and at the same numerical density as in the cloud center, $n\equiv n(0,0)$, $P_{\mathrm{ideal}}=2\pi n^2\hbar^2/m$.

The particle number $N$ is found by integrating $n_1(x)$. The numerical density profile $n(x,y)$ is fully reconstructed from the integral $n_1(x)$ owing to the cylindrical symmetry of the potential~(\ref{eq:OptLattice}) in stretched coordinates $(x,\tilde{y}\equiv y\,\omega_y/\omega_x)$. The inverse Abel transform yields
\begin{equation}
n(\tilde{\rho})=-\frac{\omega_y/\omega_x}\pi\int_{\tilde{\rho}}^\infty\frac{dn_1}{dx} \frac{\rmd x}{\sqrt{x^2-\tilde{\rho}^2}},
\label{eq:InvAbel}
\end{equation}
where $\tilde\rho\equiv\sqrt{x^2+\tilde{y}^2}$. Profile $n(\tilde\rho)$ is displayed in figure~\ref{fig:DensityProfiles}b. It is known that the inverse Abel transform emphasizes noise, especially on small scale.
To avoid the noise in the $n(\tilde\rho)$ distribution, the small-scale noise in profile $n_1(x)$ is filtered out prior to the substitution into formula (\ref{eq:InvAbel}).
The fit of the parabola $n(\tilde\rho)=n-\tilde\rho^2n''/2$ to the data near the origin yields the sought quantity $n$.

\begin{figure}[htb!]
\begin{center}
\includegraphics[width=\linewidth]{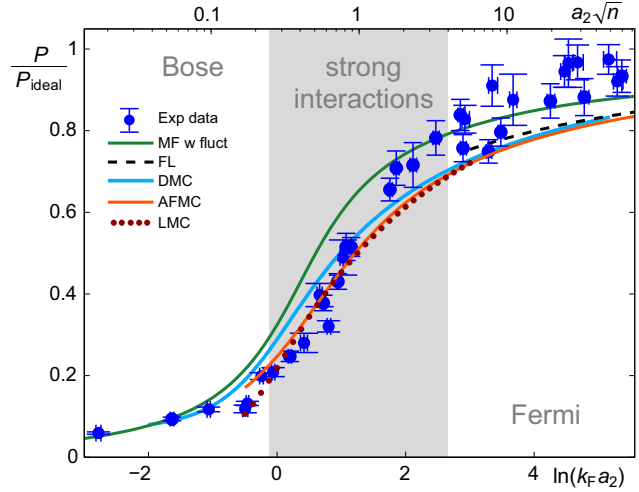}
\end{center}
\caption{Normalized pressure at the cloud center vs the interaction parameter.
Blue dots are the data~\cite{FermiBose2DCrossover}.
Green curve is the mean-field model supplemented by fluctuations~\cite{Fermi2DMeanFieldPlusFluctuations2015}.
Dashed curve is the Fermi-liquid theory~\cite{FermiLiquid2D1992}.
Light-blue solid curve is a diffusion Monte Carlo~\cite{Giorgini2D2011}.
Orange curve is an auxiliary-field Monte Carlo~\cite{Fermi2DExactGS2015}.
Burgundy dots is a lattice Monte Carlo~\cite{Fermi2DAbInitioLattice2015}.
}
\label{fig:PvskFa2}
\end{figure}

The measured normalized pressure at the cloud center, $P_2/P_{2\,\mathrm{ideal}}$, vs the interaction parameter is shown in figure~\ref{fig:PvskFa2}.
The measurement destroys the state of the system. Therefore, for each measurement the system is prepared from the beginning.

The qualitative form of the normalized pressure dependence on the coupling parameter tells that the system crosses over from fermionic statistics (in the right-hand side of figure~\ref{fig:PvskFa2}) to bosonic statistics (in the left-hand side of figure~\ref{fig:PvskFa2}). At $a_2\sqrt{n}\gg1$ the pressure value closely approaches the Fermi pressure. Such high pressure may not be explained by thermal effects since the temperature is $\ll\varepsilon_{\mathrm{F}}$.
The pressure measured in the Bose regime is due to a weak repulsion between diatomic molecular bosons. The center of the strong-interaction region $a_2\sqrt{n}=1$ approximately corresponds to the pressure drop by a factor of 2.

In the Fermi region $a_2\sqrt{n}\geq5$, the temperature is in the interval $T=(0.02$--$0.15)E_{\mathrm{F}}$. Meanwhile, at $a_2\sqrt{n}=5$ the temperature of pair breaking is expected at $0.01E_{\mathrm{F}}$~\cite{Shlyapnikov2DFermi2003}, and even lower at higher values of $a_2\sqrt{n}$. Therefore, in the Fermi regime, the superfluid phase is most probably absent and the system is in the Fermi-liquid state.

\section{Critical phenomena in the 2D Fermi-to-Bose crossover}\label{sec:CriticalPhenomena}
\subsection{BKT corrections to mean field results in dilute gases}

In the BEC domain of local pairs or molecules for the dilute 2D Fermi gas ($k_{\mathrm{F}}r_0\ll1$) the mean field $T_{\mathrm{c}}^{\mathrm{BEC}}$ according to Fisher--Hohenberg theory is only slightly different from the exact Berezinskii-Kosterlitz-Thouless critical temperature:
\begin{equation}
\frac{\left|T_{\mathrm{c}}^{\mathrm{BEC}}-T_{\mathrm{c}}^{\mathrm{BKT}}\right|}{T_{\mathrm{c}}^{\mathrm{BEC}}}\sim f_{2-2}\ll1.
\end{equation}
In the BCS domain of extended pairs in the weak-coupling case $|E_{\mathrm{b}}|\ll\varepsilon_{\mathrm{F}}$ the mean-field $T_{\mathrm{c}}^{\mathrm{BCS}}$ \cite{Orlando2DSuperconductor1979} is close again to exact critical temperature $T_{\mathrm{c}}^{\mathrm{BKT}}$:
\begin{equation}
\frac{\left|T_{\mathrm{c}}^{\mathrm{BCS}}-T_{\mathrm{c}}^{\mathrm{BKT}}\right|}{T_{\mathrm{c}}^{\mathrm{BCS}}}\sim \frac{T_{\mathrm{c}}^{\mathrm{BCS}}}{\varepsilon_{\mathrm{F}}}\sim \sqrt{\frac{|E_{\mathrm{b}}|}{\varepsilon_{\mathrm{F}}}}.
\end{equation}
For the intermediate coupling case $|E_{\mathrm{b}}|\sim\varepsilon_{\mathrm{F}}$ the difference between exact and mean-field critical temperatures becomes substantial.

Several layers or slabs of dilute gases may be created such as shown in figure~\ref{fig:TrapAnd2D}b. Therefore, possible role of the Josephson (tunneling) coupling~\cite{Josephson1961,Josephson1965} between the layers has to be considered also.
This coupling can modify the one-particle spectrum from $\varepsilon(p_\perp)=p_\perp^2/(2m)$ to $\varepsilon(p_\perp,p_z)=p_\perp^2/(2m)+J[1-\cos(p_zd)]$, where $d$ is an interlayer distance. The inclusion of $J$ rapidly makes the spectrum quasi-three-dimensional and suppresses fluctuations~\cite{MKagan1993ParallelB}. An experimental way of suppressing Josephson coupling is to make $J$ smaller than the temperature or the inversed duration of the experiment.

Note that typical for BKT transition creation and destruction of vortex-antivortex pairs as well as specific form of pair correlation function takes place close to exact critical temperature $T_{\mathrm{c}}^{\mathrm{BKT}}$, while for low temperature $T\ll T_{\mathrm{c}}^{\mathrm{BKT}}$ these features are much less pronounced and mean-field approaches are more applicable.

\subsection{Cooper-pair condensation in the strongly-interacting regime}

\begin{figure*}[htb!]
\begin{center}\includegraphics[width=0.9\linewidth]{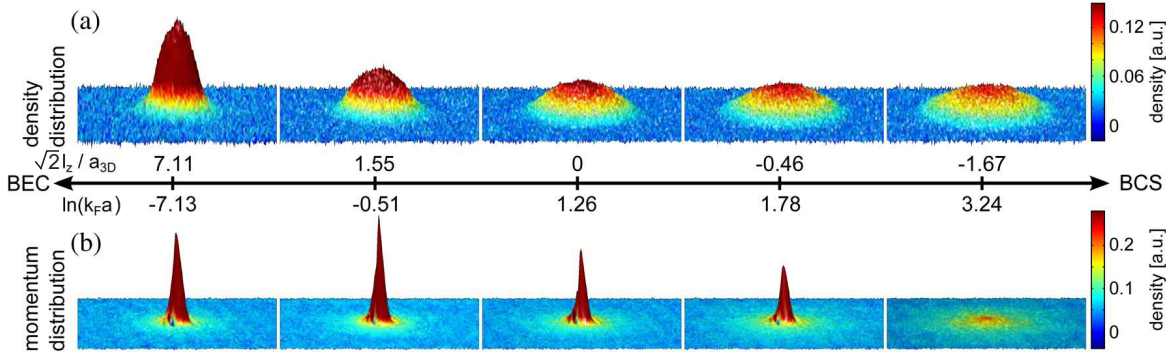}\end{center}
\caption{Density distributions at the lowest accessible temperature for different interaction strengths. (a) \textit{In situ} density distribution obtained from absorption imaging along the $z$ axis. (b) Pair momentum distribution obtained from the $\tau/4$ method with a pair projection ramp to $\sqrt2l_z/a_{\mathrm{3D}}=7.11$ (692 G). The strong enhancement at low momenta in the momentum distribution for $\ln k_{\mathrm{F}}a<3.24$ is a clear signature of pair condensation. Each picture is the average of about 30 individual measurements. The temperature of the samples ranges from 64 nK at $\ln k_{\mathrm{F}}a=-7.13$ to 78 nK at $\ln k_{\mathrm{F}}a=3.24$.
Adapted with permission from \cite{Jochim2DPairCondensation2015} copyrighted by the American Physical Society; notations altered. Also note that in \cite{Jochim2DPairCondensation2015}, $a=a_2\rme^{\gamma_{\mathrm{E}}}/2$ is calculated in the zero-momentum limit~(\ref{eq:a2atZeroEnergy}).
}\label{fig:JochimCondensation}
\end{figure*}

A signature of Cooper-pair condensation has been observed in \cite{Jochim2DPairCondensation2015}. An equally populated mixture of $^6$Li atoms in two spin states has been studied near the $B=832$~G 3D Feshbach resonance, with $N=4$--$5\times10^4$ atoms per spin state. The kinematics is restricted by confining in a disc-shaped trap with the anisotropy ratio $\omega_z/\omega_\perp\simeq310$, which gives $E_{\mathrm{F}}/(\hbar\omega_z)=0.91$--$1.02$ and, for the considered couplings, $\tilde\mu<\hbar\omega_z$. The authors have measured the 2D density distribution $n(\boldsymbol\rho)$ in the trap (figure~\ref{fig:JochimCondensation}a) and also after abrupt extinction of the potential along the $z$ direction and subsequent evolution in a radial potential (figure~\ref{fig:JochimCondensation}b).
All the measurements are performed at the lowest attainable temperatures. The in-trap images (figure~\ref{fig:JochimCondensation}a) show smooth shrinking of the gas as the atom-atom attraction is increased, while a more dramatic effect is observed in the images taken after the evolution (figure~\ref{fig:JochimCondensation}b).

The images preceded by the evolution (figure~\ref{fig:JochimCondensation}b) may be interpreted as Bose condensation of atomic pairs. Upon preparation of the gas at the reported interaction strength, the magnetic field is quickly shifted to the smallest value on the Bose side of the resonance ($B=692$ G, $\sqrt2l_z/a_{\mathrm{3D}}=7.11$). The shift is performed on time scale $\ll1/\omega_\perp$. During this shift, the radial density profile does not change, while the atoms bind into small pairs, at least if they were bound prior to the shift. Immediately after the change in the interaction strength, the optical confining potential is instantaneously extinguished, which removes the axial and weakens the radial confinement. As a result the cloud freely expands in the $z$ direction and evolves in the radial potential with the frequency $\omega_{\mathrm{exp}}/(2\pi)\simeq10$~Hz. The cloud is photographed after the evolution over the quarter of the radial period, $\tau/4\equiv\pi/(2\omega_{\mathrm{exp}})$. The images are shown in figure~\ref{fig:JochimCondensation}b. The authors suggest that the density distribution imaged after the evolution corresponds to the momentum distribution right before the release. In this case, the sharp peaks signify Bose condensation of the pairs of atoms.

The interpretation of figure~\ref{fig:JochimCondensation}b depends on the hypothesis that the expansion is collisionless. In this case the radial harmonic potential initially acts as a lens and, as a result, the density distribution after the evolution for $t=\tau/4$, $n(\boldsymbol\rho,t=\tau/4)$, coincides with the initial momentum distribution, $\tilde n(\bi{k},t=0)$, up to a constant factor. Weakening the interaction by ramping the magnetic field right before the expansion helps in reduction of the collision rate.

Superfluid hydrodynamics~\cite{StringariHydro2002,OHaraScience}, however, is an alternative scenario of the expansion. Such scenario requires nonzero superfluid order parameter $\psi$ but does not require condensation. In such expansion, nearly all momentum is released along the tightly confined direction, with almost no momentum released radially. Hydrodynamic expansion, therefore, may also produce sharp peaks similar to those of figure~\ref{fig:JochimCondensation}b. Therefore, the experiment \cite{Jochim2DPairCondensation2015} may be observing onset of superfluidity or quasicondensation rather than condensation. Same group also points out that the temperature, at which the peaks appear, coincides with the temperature of the Berezinskii--Kosterlitz--Thouless transition, which they measure~\cite{JochimBKT2015}.

Superfluid hydrodynamics is also discussed in section~\ref{sec:Criteria2D}.

\subsection{Berezinskii--Kosterlitz--Thouless transition}

Proliferation of vortices and the change in the correlation decay law are the two hallmarks of the Berezinskii--Kosterlitz--Thouless transition~\cite{BerezinskiiIIeng,KosterlitzThouless1973} from a superfluid to normal state.
While in a weakly-interacting Bose gases the appearance~\cite{DalibardBKT} and pairing~\cite{VortexPairing2013} of vortices has been detected, such effects have not been observed in the Bose-to-Fermi crossover, where the interactions between Bose molecules are typically stronger than in the atomic Bose gases.

\begin{figure}[htb!]
\begin{center}
\includegraphics[width=0.9\linewidth]{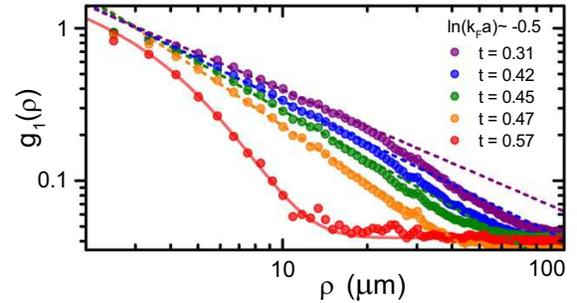}
\end{center}
\caption{First-order correlation function $g_1(\rho)$ for different temperatures at $\ln(k_{\mathrm{F}}a)=-0.5$. The temperature scale used here is $t=T/T^0_{\mathrm{BEC}}$. At high temperatures, correlations decay exponentially as expected for a gas in the normal phase. At low temperatures, algebraic correlations [$g_1(\rho)\propto\rho^{-\eta(T)}$] with a temperature-dependent scaling exponent $\eta(T)$ are observed.
Adapted with permission from \cite{JochimBKT2015} copyrighted by the American Physical Society; notations altered. Also note that in \cite{JochimBKT2015}, $a=a_2\rme^{\gamma_{\mathrm{E}}}/2$ is calculated in the zero-momentum limit~(\ref{eq:a2atZeroEnergy}).}\label{fig:BKTg1}
\end{figure}

Switching from a power-law to exponential decay of the phase correlation in response to the rising temperature has been recently observed~\cite{JochimBKT2015} in the Bose-to-Fermi crossover. The experimental setup and techniques in many respects are the same as in \cite{Jochim2DPairCondensation2015}, which is discussed in the pervious section. The first order correlation function $g_1(\boldsymbol\rho)$ is derived from the measured momentum distribution $\tilde n(\bi{k})$:
\begin{equation}
g_1(\boldsymbol\rho)=\int\tilde n(\bi{k})e^{i\bi{k}\cdot\boldsymbol\rho}\rmd^2k.
\end{equation}
The momentum distribution $\tilde n(\bi{k})$ is obtained by the focusing technique as explained in the previous section. The result of such measurement of the correlation function is shown in figure~\ref{fig:BKTg1}.
All the measurements reflected in this figure are performed at the same interaction strength, $\ln(k_{\mathrm{F}}a)\simeq-0.5$, belonging to the strongly interacting regime between the Bose and Fermi asymptotes. The only parameter, which is varied in the experiment, is the dimensionless temperature $t=T/T^0_{\mathrm{BEC}}$, where $T^0_{\mathrm{BEC}}=E_{\mathrm{F}}\sqrt3/\pi$ is the ideal-gas Bose-condensation temperature. One may see that at lower temperatures the decay is algebraic [$g_1(r)\propto r^{-\eta(T)}$] with a temperature-dependent exponent $\eta(T)$, while in a hotter gas the correlation function decays exponentially. The transition from exponential to power low decay signals the onset of superfluidity. Interestingly, the observed value of the exponent $\eta=0.6$--$1.4$ is well above the values $\eta\leq0.25$ expected for a homogeneous gas. Inhomogeneity may indeed cause an increase of $\eta$ compared to uniform gases as found within quantum-Monte-Carlo simulation~\cite{JochimBKT2015}.

The measurement of the correlation decay rests upon the assumption of the ballistic cloud expansion after abrupt extinction of the potential along $z$. In the superfluid phase, the expansion may be going along an alternative scenario of superfluid hydrodynamics. This would complicate the relation between $n(\boldsymbol\rho,\tau/4)$ and the sought $\tilde n(\bi{k},0)$ as well as the derivation of $g_1(\boldsymbol\rho)$ from the data.
In the normal phase, however, the expansion should be nearly ballistic due to a low collision number. Therefore, the power-law decay of the measured $g_1(\boldsymbol\rho)$ cannot be obtained in a normal gas and does indeed indicate the onset of superfluidity.

Recent calculations of the algebraic-decay exponent have accounted for the trapped-gas inhomogeneity \cite{LevinBKT2D2015,BoettcherBKTExponent2016} and contribution of the normal fraction~\cite{BoettcherBKTExponent2016}. These calculations have shown $\eta$ values close to the measured ones \cite{BoettcherBKTExponent2016}. This agreement may be regarded as evidence against hydrodynamic expansion because such expansion would have changed the apparent $\eta$ value, making it different from the actual \textit{in situ} quantity.

\section{Equation of state in the crossover}\label{sec:Models}
\subsection{The pressure in 2D Fermi-to-Bose crossover}

At low temperatures the pressure $P=2n\mu-E$, where we used thermodynamic identity $\mu=(1/2)\partial E/\partial n$ (note that the total density is $2n$). Let us consider first the BCS case and the intermediate case. Taking into account that in 2D gas with parabolic spectrum $\varepsilon_{\mathrm{F}}=2\pi n/m$ and total energy density (energy per unit surface) $E=n\varepsilon_{\mathrm{F}}-n|E_{\mathrm{b}}|$, we get for the total pressure at the mean-field level:
\begin{equation}
P\simeq\frac{n\varepsilon_{\mathrm{F}}}2.
\label{eq:IdealFermiPressure2D}
\end{equation}
It is a remarkable result which contains only the Pauli pressure.

Note that, the Fermi-gas expansion in the normal (nonsuperfluid) state both for the total energy and pressure gives the corrections proportional to $f_0(k_{\mathrm{F}}a)$ and $f_0^2(k_{\mathrm{F}}a)$, where in terms of density per spin state $n$:
\begin{equation}
k_{\mathrm{F}}=\sqrt{4\pi n}.
\end{equation}
Thus deep in the BCS domain instead of a simple expression (\ref{eq:IdealFermiPressure2D}) for the pressure we should have:
\begin{equation}
P=n\varepsilon_{\mathrm{F}}\left[1+c_1f_0(k_{\mathrm{F}}a)+ c_2f_0^2(k_{\mathrm{F}}a)\right],
\label{eq:MeanFieldPressure2D}
\end{equation}
where $c_1$ and $c_2$ are numerical constants. Note that first-order corrections in $f_0$ to the expression (\ref{eq:IdealFermiPressure2D}) were obtained in \cite{HardCore2DBosons1971}.

At the same time deep in the BEC domain the total energy is given by:
\begin{equation}
E\simeq-n|E_{\mathrm{b}}|+\frac{\pi n^2}mf_{2-2},
\end{equation}
where we again neglect the small corrections of the order of $\varepsilon_{\mathrm{F}}/W$ and $W/|E_{\mathrm{b}}|$. Hence for the pressure we get
\begin{equation}
P\simeq\frac{\pi n^2}mf_{2-2}.
\label{eq:MeanFieldBosePressure2D}
\end{equation}

A dilute two-dimensional Fermi gas with \textit{s}-wave interaction is one of the few cases where the Fermi-liquid parameters were calculated from first principles~\cite{Bloom1975,FermiLiquid2D1992}. Using the results of \cite{FermiLiquid2D1992} one may find the pressure of the interacting normal (nonsuperfluid) Fermi gas
up to the $f_0^2$ term:
\begin{equation}
\frac{P}{P_{\mathrm{ideal}}}=1+2f_0+(7-4\ln2)f_0^2,
\label{eq:Pressure2DFermiLiquid}
\end{equation}
where $f_0$ is given by (\ref{eq:GasParameter2D}), while $P_{\mathrm{ideal}}=2\pi\hbar^2n^2/m$ and coincides with (\ref{eq:IdealFermiPressure2D}).

\subsection{Compressibility and sound velocity in the 2D BCS-BEC crossover}

Evaluation of the pressure in the 2D BCS-BEC crossover allows us to calculate the speed of sound $c$ (or equivalently the inverse compressibility $\kappa^{-1}\sim c^2$). Using the well-known thermodynamic identities $c^2\sim\partial P/\partial\rho$, where $\rho=2nm$ is the mass density, and $dP=2n\,d\mu$, we get at low temperatures:
\begin{equation}
c^2=\frac1{2m}\frac{\partial P}{\partial n}=\frac nm\frac{\partial\mu}{\partial n},
\end{equation}
where $\mu$ is the chemical potential. Thus from the expression (\ref{eq:MeanFieldPressure2D}) for the pressure in the BCS domain we get for the sound velocity:
\begin{equation}
c^2\simeq\frac{v_{\mathrm{F}}^2}2\left[1+c_1f_0+c_3f_0^2\right],
\end{equation}
where $v_{\mathrm{F}}=\hbar k_{\mathrm{F}}/m$ is the Fermi velocity. Analogously deep in the BEC domain from (\ref{eq:MeanFieldBosePressure2D}) we have:
\begin{equation}
c^2\simeq\frac{v_{\mathrm{F}}^2}4f_{2-2}\left[1+\mathcal{O}(f_{2-2})\right].
\end{equation}
It is important to emphasize that the negative values of the sound velocity squared (or of the inverse compressibility) signal the instability of the ground state towards some kind of the phase separation (see, for example, \cite{MKaganPhaseSeparation2001,MKaganPolarons2011,MKaganPolarons2017}).

\subsection{Scale invariance and scale anomaly}

The Hamiltonian of a 2D uniform gas with pair interactions may generally be taken in the form
\begin{equation}
\hat H=\sum_i-\frac{\hbar^2}{2m}\Delta_i+\sum_{i<j}V(\boldsymbol{\rho}_i-\boldsymbol{\rho}_j).
\end{equation}
In case of the contact potential $V(\boldsymbol{\rho'})\propto\delta^2(\boldsymbol{\rho'})$ the Hamiltonian is invariant under the dilatation transformation $\boldsymbol{\rho}_i\rightarrow\lambda\boldsymbol{\rho}_i$ \cite{PitaevskiiScaleInvariance2D}. Such transformation only changes the Hamiltonian as $\hat H\rightarrow \hat H/\lambda^2$, which implies the scale invariance of the solutions.

An external trapping potential breaks the scale invariance. For symmetric trap $m\omega^2\rho^2/2$, however, the system remains symmetric with the SO(2,1) symmetry~\cite{PitaevskiiScaleInvariance2D}. Such symmetry uniquely defines the frequency of the breathing mode in the harmonic trap. Regardless of the interaction strength or particle statistics, the mode frequency is $2\omega$.
For the zero-temperature gas the the local equation of state $\tilde\mu\propto n$ and $P\propto n\varepsilon_{\mathrm{F}}$ is expected~\cite{HofmannScaleAnomaly2012,RanderiaScaleInv2012}.

\begin{figure}[htb!]
\begin{center}\includegraphics[width=0.9\linewidth]{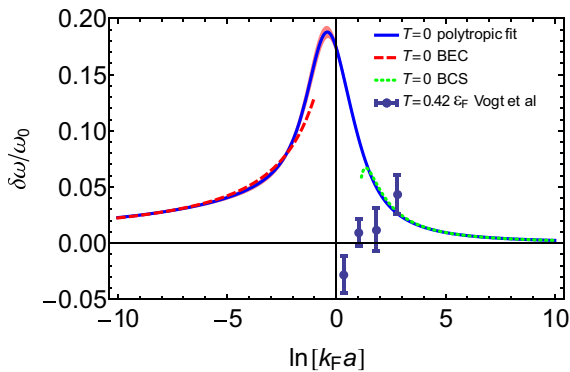}\end{center}
\caption{Anomalous frequency shift $\delta\omega$ at $T=0$ as determined from the fit to the equation of state in \cite{Giorgini2D2011}. The red error band indicates the propagated error from the Monte Carlo calculation. The plot also shows the exact results in the BCS (green dotted line) and BEC limit (red dashed line). For comparison, experimental values are included for weak breathing mode excitations at $T=0.42\varepsilon_{\mathrm{F}}$ as reported in \cite{Kohl2DBreathingMode2012}. It remains an open question how thermal fluctuations affect the zero temperature results.
Adapted with permission from \cite{HofmannScaleAnomaly2012} copyrighted by the American Physical Society; notations altered.
}\label{fig:ScaleAnomHofmann}
\end{figure}

The scale invariance, however, may be broken by regularization of the $\delta^2(\boldsymbol{\rho'})$ potential~\cite{PitaevskiiScaleInvariance2D}.
The regularization also breaks the SO(2,1) symmetry and the invariance of the breathing mode frequency. On the basis of Monte-Carlo simulation of the equation of state~\cite{Giorgini2D2011}, the frequency of the breathing mode has been calculated~\cite{HofmannScaleAnomaly2012}. The amount of the scale-anomalous shift of the breathing mode frequency is shown in figure~\ref{fig:ScaleAnomHofmann}.
The value of the shift has also been calculated in \cite{ScaleInvBreakingGaoYu2012,ScaleInvSchaefer2013}.
Since the anomaly is related to regularization of the interactions, the shift is the largest where the interactions are the strongest. The predicted shift is large enough to be detectable in ultracold-atom experiments.

The measurement of the breathing mode frequency~\cite{Kohl2DBreathingMode2012} did not show any obvious shift with respect to the scale-invariant prediction $2\omega$. The conditions for the calculations and the experiment differ in two respects. Firstly, the equation of state has been calculated for zero temperature ($T=0$) \cite{Giorgini2D2011}, while the temperature $T=0.42\varepsilon_{\mathrm{F}}$ is reported in the experiment. The frequency shift at $T=0$ is the upper bound for the finite-$T$ shift~\cite{RanderiaScaleInv2012}; the lower bound, however, is unknown. Secondly, the shift is calculated for a purely 2D atom-atom interaction potential, while the atoms interact via a 3D potential. While 2D and 3D scattering parameters are expressed via each other as discussed in Sec.~\ref{sec:Scattering2D}, the regularization procedure and related anomalous shift may be different from the purely 2D case.

\subsection{Testing zero-temperature models}

\begin{figure}[htb!]
\begin{center}\includegraphics[width=\linewidth]{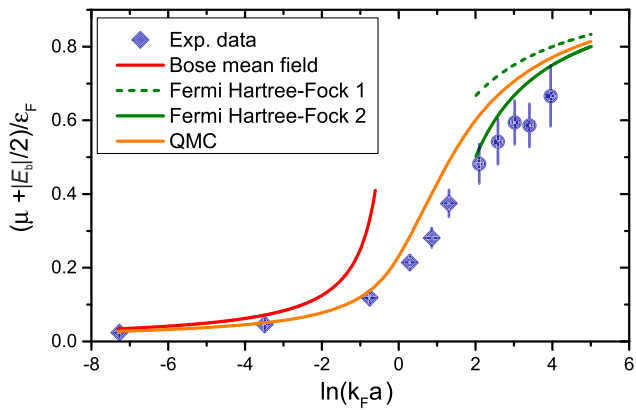}\end{center}
\caption{Low-temperature equation of state across the 2D BEC-BCS crossover. The experimental results are obtained from measurements of the quasi-2D gas at the lowest attainable temperatures, which corresponds to $T/\varepsilon_{\mathrm{F}}\simeq0.05$ and $0.1$ on the Bose and Fermi sides. The data points shown as diamonds (circles) correspond to measurements in the superfluid (normal) phase. The solid red line on the Bose side corresponds to the mean field formula $\tilde\mu/\varepsilon_{\mathrm{F}}=-1/[4\ln(k_{\mathrm{F}}a)]$, whereas the dashed and solid green lines on the Fermi side display the non-self-consistent and self-consistent Hartree--Fock predictions $1/[1+\ln^{-1}(k_{\mathrm{F}}a)]$ and $1-\ln^{-1}(k_{\mathrm{F}}a)$ for weakly attractive fermions. The orange line is the prediction for the ground state equation of state from recent auxiliary-field quantum Monte Carlo calculations~\cite{Fermi2DExactGS2015}.
Adapted with permission from \cite{Jochim2DThermodynamics2015} copyrighted by the American Physical Society; notations altered.
}\label{fig:JochimZeropTempMu}
\end{figure}

Available zero-temperature equations of state may be tested using the measurements of the pressure~\cite{FermiBose2DCrossover} shown in figure~\ref{fig:PvskFa2} and the measurements of the chemical potential~\cite{Jochim2DThermodynamics2015} displayed in figure~\ref{fig:JochimZeropTempMu}.

The analytic models for the zero-temperature gas at the Fermi-to-Bose crossover are presented by (\ref{eq:Pressure2DFermiLiquid}) and (\ref{eq:MeanFieldBosePressure2D}), for the fermionic and bosonic asymptotes respectively. Formula (\ref{eq:Pressure2DFermiLiquid}) describes the gas within the Fermi-liquid model~\cite{FermiLiquid2D1992}. Strictly speaking, the pairing (superfluid) gap is present at $T=0$. The gap, however, does not alter the pressure value significantly since the gap is small and the pressure is continuous at the transition. In figure~\ref{fig:PvskFa2}, the Fermi-liquid prediction is shown by the dashed curve. This analytical result agrees with the fixed-node diffusion Monte Carlo numerical calculation~\cite{Giorgini2D2011} and auxiliary-field Monte Carlo~\cite{Fermi2DExactGS2015}, but differs from the measurements. The discrepancy with the measurement is significant because within the Fermi-liquid theory, the pressure is counted from the Fermi pressure rather than from zero. The difference between the experimental data and the Fermi-liquid-theory~\cite{FermiLiquid2D1992} may be due to the mesoscopic character of the experimental system~\cite{FermiBose2DCrossover} since the calculation is done for infinitely extended Fermi liquid.

In the Fermi regime, the data of figure~\ref{fig:JochimZeropTempMu} are in contradiction with the data of figure~\ref{fig:PvskFa2} discussed above.
In~\ref{fig:JochimZeropTempMu}, the data are consistently below the auxiliary-field Monte Carlo~\cite{Fermi2DExactGS2015} (orange curve), while the data of figure~\ref{fig:PvskFa2} are consistently above.

For a 3D gas, the analytic mean-field approach of Leggett's type \cite{BlochLowDReview2008,MKaganCrossover2006} qualitatively correctly predicts the pressure and chemical potential at any interaction level. Upon inclusion of fluctuations, this model comes in reasonable agreement with the experiments, including the Bose regime, where original mean-field model has the largest discrepancy.

For a 2D system, the use of Leggett's type approach is limited. While a two-body quantity (the two-fermion binding energy) shows qualitatively correct behavior increasing with the interaction growth~\cite{Randeria2DCrossover1989prl}, many-body quantities such as the pressure, on the contrary, occurred to be independent of the interaction. In particular, such model predicts that the Fermi pressure must be present in the Bose limit, at least in the range $|E_{\mathrm{b}}|\sim\varepsilon_{\mathrm{F}}$. The addition of fluctuations into the mean-field model yields qualitatively correct dependence of the pressure on the interaction~\cite{Fermi2DMeanFieldPlusFluctuations2015}, shown by the green curve in figure~\ref{fig:PvskFa2}. Indeed, according to the Ginzburg--Levanyuk criterion~\cite{GinzburgLevanyuk1959eng,GinzburgLevanyuk1960eng}, the reduction of the dimensionality is accompanied by the increasing role of fluctuations (see also \cite{LoktevReview2001}).

Models with qualitatively correct description of the Fermi-to-Bose crossover appeared only in the recent few years, such as the model~\cite{Fermi2DMeanFieldPlusFluctuations2015}. These models include quantum diffusion Monte Carlo simulations~\cite{Giorgini2D2011,DiffusionMC2015}, self-consistent $T$-matrix~\cite{Fermi2DEOSandPressureParish2014} (see also earlier papers \cite{Miyake1983,MKaganSpectralFunc1998,MKaganSpectralFunc2000}), finite-temperature lattice Monte Carlo~\cite{Fermi2DAbInitioLattice2015}, and auxiliary-field Monte Carlo~\cite{Fermi2DExactGS2015}. Predictions of some of these models are shown in figure~\ref{fig:PvskFa2}, while the latter model is also displayed in figure~\ref{fig:JochimZeropTempMu}. Results of all the models are given for $T=0$.

\begin{figure}[htb!]
\begin{center}
\includegraphics[width=0.9\linewidth]{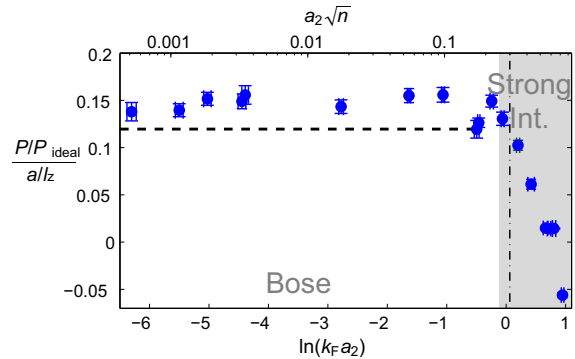}
\end{center}
\caption{Demonstration of linear scaling $P/P_{\mathrm{ideal}}\propto a/l_z$ in the Bose regime of the Fermi-to-Bose crossover. Markers: The data $(P/P_{\mathrm{ideal}})/(a/l_z)$ vs the interaction parameter~\cite{FermiBose2DCrossover}. Dashed horizontal line: Model for pointlike molecular bosons with 3D interactions, $P=P_{\mathrm{ideal}}\frac{0.6a_{\mathrm{3D}}}{l_z\sqrt{8\pi}}$.}
\label{fig:BoseLinear}
\end{figure}

For the Bose regime, the analytic prediction is given by (\ref{eq:MeanFieldBosePressure2D}). The leading term in the pressure of Bose molecules should be the same as for point-like bosons: $P_{\mathrm{Bose}}=-P_{\mathrm{ideal}}/[4\ln(k_{\mathrm{F}}a_{2-2})]$~\cite{HardCore2DBosons1971}. Here $a_{2-2}$ is the 2D scattering length for molecule-molecule collisions, which may be related to the respective 3D scattering length $a_{2-2}^{\mathrm{3D}}=0.6\,a^{\mathrm{3D}}$~\cite{Petrov,MKagan06a2005,MKagan06a2006} by equating the scattering amplitudes $f_{\mathrm{2D}}(2q\rightarrow0,a_{2-2})= f_{\mathrm{Q2D}}(2q\rightarrow0,a_{2-2}^{\mathrm{3D}},l_z/\sqrt2)$.
The low-energy limit $2\tilde\mu\ll\hbar\omega_z$ yields
\begin{equation}
a_{2-2}\simeq2.09\,l_z\exp\left(-\sqrt{\frac\pi2}\frac{l_z}{a_{2-2}^{\mathrm{3D}}}\right).
\end{equation}
As a result, in the limit of unmodified 3D interactions $a_{2-2}^{\mathrm{3D}}\ll l_z$, one obtains
\begin{equation}
P_{\mathrm{Bose}}\simeq P_{\mathrm{ideal}}\frac{a_{2-2}^{\mathrm{3D}}}{l_z\sqrt{8\pi}},
\end{equation}
which is shown by the dashed line in figure~\ref{fig:BoseLinear}.
We see that in the Bose regime $P/P_{\mathrm{ideal}}$ scales as $a/l_z$. This scaling is applicable in a wide range.

\subsection{Finite-temperature equations of state}

Recent progress in virial thermometry \cite{Vale2DThermodynamics2015,Jochim2DThermodynamics2015}, discussed in section~\ref{sec:Thermometry}, made it possible to take data for testing finite temperature equations of state. The thermometry has been accompanied by the measurement of the chemical potential $\mu$. As a result, a measured local density can be expressed as a function of $\mu$ and $T$. A related dimensionless value, namely phase space density $n\lambda_{\mathrm{dB}}^2$, may be expressed as a function of dimensionless variables $\mu/T$ and $|E_{\mathrm{b}}|/T$~\cite{Vale2DThermodynamics2015}. Another convenient dimensionless value is $n/n_0$, where $n_0$ is the local density of an ideal Fermi gas at the same values of $\mu$ and $T$:
\begin{equation}
n_0=\frac1{\lambda_{\mathrm{dB}}^2}\ln(1+e^{\mu/T}).
\end{equation}
Combination $n/n_0$ also includes the phase space density times a function of fugacity.

\begin{figure}[htb!]
\begin{center}
\includegraphics[width=\linewidth]{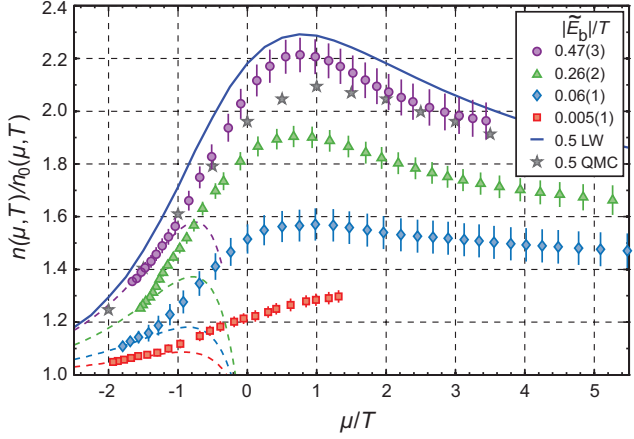}
\end{center}
\caption{Normalized density equation of state for a 2D attractive Fermi gas for $|\tilde E_{\mathrm{b}}|/T=0.47(3)$ (purple circles), $|\tilde E_{\mathrm{b}}|/T=0.26(2)$ (green triangles), $|\tilde E_{\mathrm{b}}|/T=0.06(1)$ (blue diamonds), and $|\tilde E_{\mathrm{b}}|/T=0.005(1)$ (red squares). Dashed lines show the calculated equation of state using the third order virial expansion~\cite{DrummondVirial2D2010}. Solid blue line and grey stars are the calculated equation of state for $|\tilde E_{\mathrm{b}}|/T=0.5$ based on Luttinger--Ward (LW) \cite{Fermi2DEOSandPressureParish2014} and lattice quantum Monte Carlo (QMC) calculations~\cite{Fermi2DAbInitioLattice2015}, respectively.
Adapted with permission from \cite{Vale2DThermodynamics2015} copyrighted by the American Physical Society; notations altered.
}\label{fig:ValeEOS}
\end{figure}

The normalized density, $n/n_0$, is represented in figure~\ref{fig:ValeEOS}.
The logarithm of the local fugacity is taken as the horizontal axis.
Interestingly, the graph is non-monotonic, which is qualitatively different from the similar measurements in 3D \cite{Vale2DThermodynamics2015}.
The authors mention that data are taken above the superfluid transition. Therefore, no superfluidity-related features are expected to be seen in figure~\ref{fig:ValeEOS}. This measurement is reported for the fermionic and strongly-interacting regime~\cite{Vale2DThermodynamics2015}.

\begin{figure*}[htb!]
\begin{center}
\includegraphics[width=\linewidth]{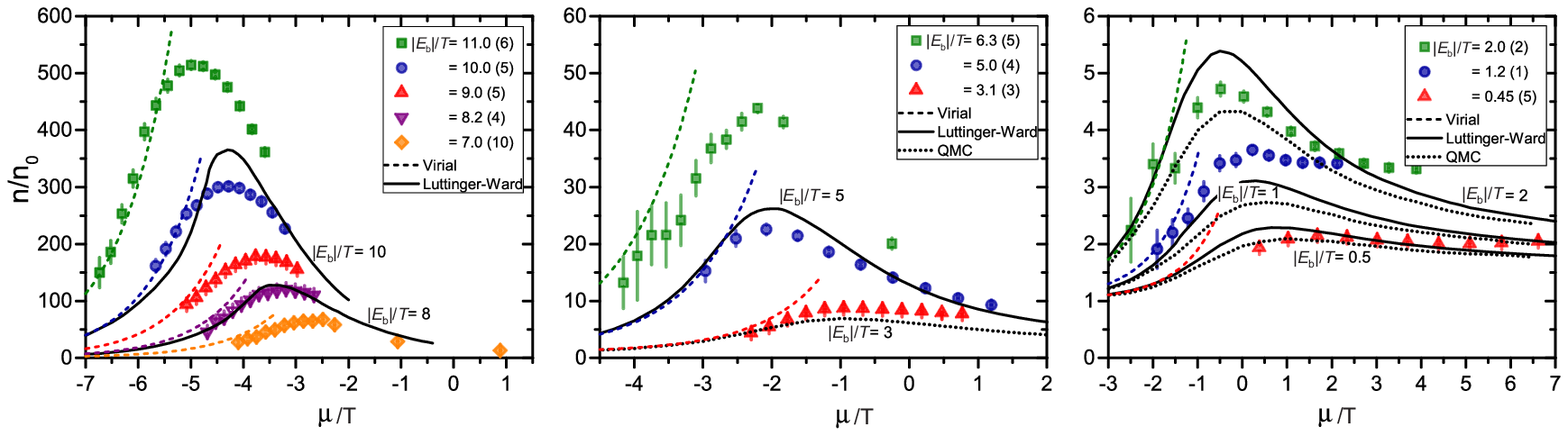}
\end{center}
\caption{EOS in the crossover regime shown as the density normalized by the ideal Fermi gas $n_0(\mu,T)=\lambda_{\mathrm{dB}}^{-2}\ln(1+e^{\mu/T})$. The experimental data points (filled shapes) are compared with the second order virial expansion at low values of $\mu/T$ (coloured dashed lines). The displayed errors are purely statistical, with systematic uncertainties estimated at 13\%--15\% and given for each data set in Table IV in the Supplemental Material of \cite{Jochim2DThermodynamics2015}. The results are compared with theoretical predictions in the 2D BEC-BCS crossover from Luttinger-Ward theory (\cite{Fermi2DEOSandPressureParish2014}, solid black lines) and fermionic QMC simulations (\cite{Fermi2DAbInitioLattice2015}, dotted black lines), with the corresponding values of $|E_{\mathrm{b}}|/T$ labeling each curve. Note that the vertical scale differs by a factor of 10 in each panel.
Adapted with permission from \cite{Jochim2DThermodynamics2015} copyrighted by the American Physical Society; notations altered.
}\label{fig:JochimEOS}
\end{figure*}

The measurements~\cite{Jochim2DThermodynamics2015} of the same quantities is reported in figure~\ref{fig:JochimEOS}.
Here the measurements cover all interaction regimes, including that where the system is mostly bosonic. In both papers~\cite{Vale2DThermodynamics2015,Jochim2DThermodynamics2015}, the data are compared to the Luttinger--Ward~\cite{Fermi2DEOSandPressureParish2014} and lattice quantum Monte Carlo~\cite{Fermi2DAbInitioLattice2015} calculations. These two models are known to make the predictions that differ in the strongly-interacting regime but agree at the other couplings.

\section{Criteria of two-dimensionality revisited}\label{sec:Criteria2D}

For comparison with 2D models, it is important that the experimental system remains two-dimensional. In this section, the effect of the interactions on the kinematic dimensionality is discussed.

Even a small two-body interaction mixes states of the $z$ motion of an atom.
From the standpoint of a single atom, therefore, even small two-body interaction brings about some deviation from the 2D kinematics. In the absence of many-body interactions, however, the center-of-mass kinematics of the pair is 2D since the confining potential is close to harmonic.
The whole system, therefore, remains kinematically 2D despite any level of two-body interactions.
For example, the bosonic asymptote of a Fermi-to-Bose crossover is a 2D gas of molecular dimers, while the motion of an atom within a dimer is 3D and may be expanded as a superposition of many eigenstates of the potential $m\omega_z^2z^2/2$ in the laboratory reference frame.

In the strongly-interacting regime, at $a_2\sqrt{n}\sim1$, the interaction is necessarily of a many-body type. Whether such interaction breaks the 2D kinematics is an important and yet unresolved question.

It has been argued that strong interactions easily break down the 2D kinematics~\cite{Vale2DCriteria2016}, \textit{e.~g.}, at $l_z/a_{\mathrm{3D}}=-0.34$, the two-dimensionality is broken at any $E_{\mathrm{F}}/(\hbar\omega_z)>0.59$.
Conclusions of \cite{Vale2DCriteria2016} are opposed below.
The arguments of \cite{Vale2DCriteria2016} are based on analyzing the disc-shaped-cloud expansion after the removal of the tight confinement. The observed expansion law changes sharply in response to increasing the atom number and, above the threshold, is faster than predicted by the noninteracting-gas model and a collisional-gas model. This change in dynamics is interpreted as a breakdown of two-dimensionality~\cite{Vale2DCriteria2016}.
As an alternative to the deviation from the 2D kinematics one may consider the onset of superfluidity in the 2D system. The onset of superfluidity is plausible for two reasons: (i) a superfluid gas may expand faster than a normal collisional gas and (ii) a sharp change is possible at the onset of superfluidity; for example, abrupt change in dynamics has been observed in the 3D Fermi gases~\cite{GrimmBreathingMode2004,HydroBreakdown}.
The authors of \cite{Vale2DCriteria2016} do not rule out superfluidity as an explanation of their very interesting data.

\begin{figure}[htb!]
\begin{center}\includegraphics[width=0.8\linewidth]{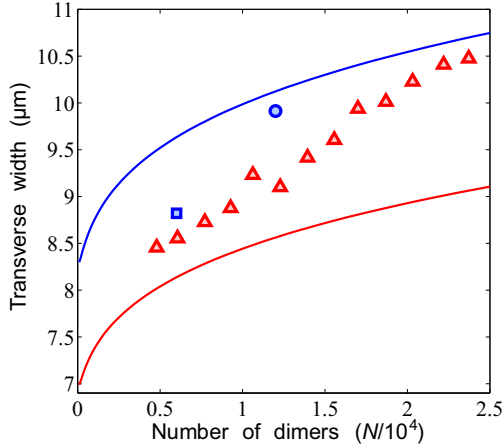}\end{center}
\caption{Expansion of 2D gas released into the 3D space by removing the $z$ confinement. The size $\Delta z$ at fixed expansion time vs $N$, the number of atoms per spin state. Symbols: data~\cite{Vale2DCriteria2016}. Curves: superfluid expansion model (\ref{eq:UnitaryGP}). Red curve: initially 2D gas with atomic width. Blue solid curve: initially 2D gas with molecular width.}\label{fig:ModelDykeFig3a}
\end{figure}

Superfluid expansion of a 2D gas released from a trap into the 3D space, described within a simple model \cite{TurlapovToPublish}, may be compared to the data of \cite{Vale2DCriteria2016} taken at the 3D Feshbach resonance, $a_{\mathrm{3D}}=\infty$. For modeling, it is assumed that immediately after the removal of the trap along $z$, the gas is a BEC of pairs with the wave function $\psi(x,y,z,t)$. Since the experiment is performed at the lowest achievable temperature, the equation of state may be taken in the form $\tilde\mu=(1+\beta)\varepsilon_{\mathrm{F}}^{\mathrm{3D}}$~\cite{OHaraScience}, where $\varepsilon_{\mathrm{F}}^{\mathrm{3D}}$ is the 3D local Fermi energy and $\beta=-0.63$ \cite{ZwierleinLambda2012,JochimNewLiFeshbach2013}. Upon the release, the dynamics of $\psi$ is assumed to obey a Gross-Pitaevskii-like equation modified for the on-resonance interactions:
\begin{eqnarray}
&&\rmi\hbar\frac{\partial\psi}{\partial t}= -\frac{\hbar^2}{2(2m)}\Delta\psi+ 2(1+\beta)\frac{\hbar^2}{2m}(6\pi^2n_{\mathrm{s}}^{\mathrm{3D}})^{2/3}\psi\nonumber\\
&&+ \frac{2m\omega_\perp^2(x^2+y^2)}2\psi,
\label{eq:UnitaryGP}
\end{eqnarray}
where $n_{\mathrm{s}}^{\mathrm{3D}}=|\psi|^2$ is the superfluid 3D numerical density per spin state. Within the model, the size along the $z$ direction ($\Delta z$) is calculated for fixed expansion time $t=600$~$\mu$s [$\omega_z/(2\pi)=5.15$~kHz and $\omega_\perp/(2\pi)=26.6$~Hz] for different $N$, number of atoms per spin state.
The result is shown in figure~\ref{fig:ModelDykeFig3a} together with the data of \cite{Vale2DCriteria2016}.
For the initial condition, the gas is assumed kinematically 2D right before the release, with the wavefunction either $\psi_0(x,y,z)\propto\exp(-z^2/(4l_z^2))$ (red solid curve) or $\psi_0(x,y,z)\propto\exp(-z^2/(2l_z^2))$ (blue solid curve). These two wavefunctions differ by the width in the $z$ direction. The first one has the width which is the same as for a point-like atom sitting in the lowest $z$ level. The second wavefunction has the same width as for a point-like diatomic molecule.
The data are between these two simple models. The data might be going towards the upper (blue) curve for two reasons: (i) the superfluid fraction increases and/or (ii) the dimension along $z$ narrows down due to a many-body contribution to the pairing energy.

Closeness of the expansion data to the results of the model suggests that the superfluid expansion scenario is plausible. This leaves possibility to quantitatively explain observations of \cite{Vale2DCriteria2016} without breakdown of two-dimensionality.

The pressure may be used as an indicator of whether the 2D kinematics is broken. Population of excited states of motion along $z$ brings about drop in $P/P_{\mathrm{ideal}}$. In figure~\ref{fig:PvskFa2}, the lowest lying models are the auxiliary-field quantum Monte Carlo~\cite{Fermi2DExactGS2015} and lattice Monte Carlo~\cite{Fermi2DAbInitioLattice2015}. The data are somewhat below these calculations in leftward side of the strongly-interacting region ($-1<\ln(k_{\mathrm{F}}a)<1$), which admits possibility of some deviation from the two-dimensionality. There are no regions, however, where the pressure is largely below available calculations.

\section{Spin-imbalanced Fermi gases}\label{sec:SpinImbal}
\subsection{Experimental progress}

Studies of 2D Fermi gases with imbalanced populations of the two spin states have been recently started~\cite{ThomasSpinImbalancedQ2D2014,SpinImbalanced2D2016}. A two-component quasi-2D gas has been prepared in harmonic traps~\cite{ThomasSpinImbalancedQ2D2014} with the different, well controlled populations, $N_\uparrow$ and $N_\downarrow$, of the two spin states.
In this system, several interesting effects may be observed.
In a general picture of fermionic systems, the experiment may have physics similar to layered electronic systems with in-plane magnetic field, where the Landau diamagnetism is absent. A strong enhancement of the critical temperature has been predicted~\cite{MKagan1993ParallelB} in such materials.
Prospectively, 2D spin-imbalanced systems are home for yet unobserved new superfluid/superconductive phases. One such phase is the Fulde--Ferell--Larkin--Ovchinnikov superfluid~\cite{FuldeFerrell,LarkinOvchinnikov} where the fermions pair and condense into states with finite momenta of the pair. In 3D the parameter range for such phase is vanishingly small~\cite{NarrowFFLO2006}, while in 2D the effect may be less fragile~\cite{FFLO2D}.
Phases with the Kohn--Luttinger pairing~\cite{KohnLuttingerPPairing1965,FayPPairing1968,MKaganPPairing1988} is another interesting example. In this case, the reduction of dimensionality down to 2D together with the spin imbalance strongly increases the critical temperature~\cite{MKagan1989KohnLuttinger}. Ultimately, the ultracold atoms may be the best system for detecting such effects due to high tunability, purity and various techniques of direct observations.

Nonuniform systems, such as quantum gases in parabolic traps, are prone to separation into a paired phase and a fully polarized phase. In 3D and 1D, this phenomenon is qualitatively different: In the 3D gases, the paired phase is in the trap center~\cite{KetterlePhaseSeparation}, while in a 1D gas, the center is occupied by a polarized fraction, while the paired atoms are pushed to the edges~\cite{Hulet1D2010}.

\begin{figure}[htb!]
\begin{center}
\includegraphics[width=\linewidth]{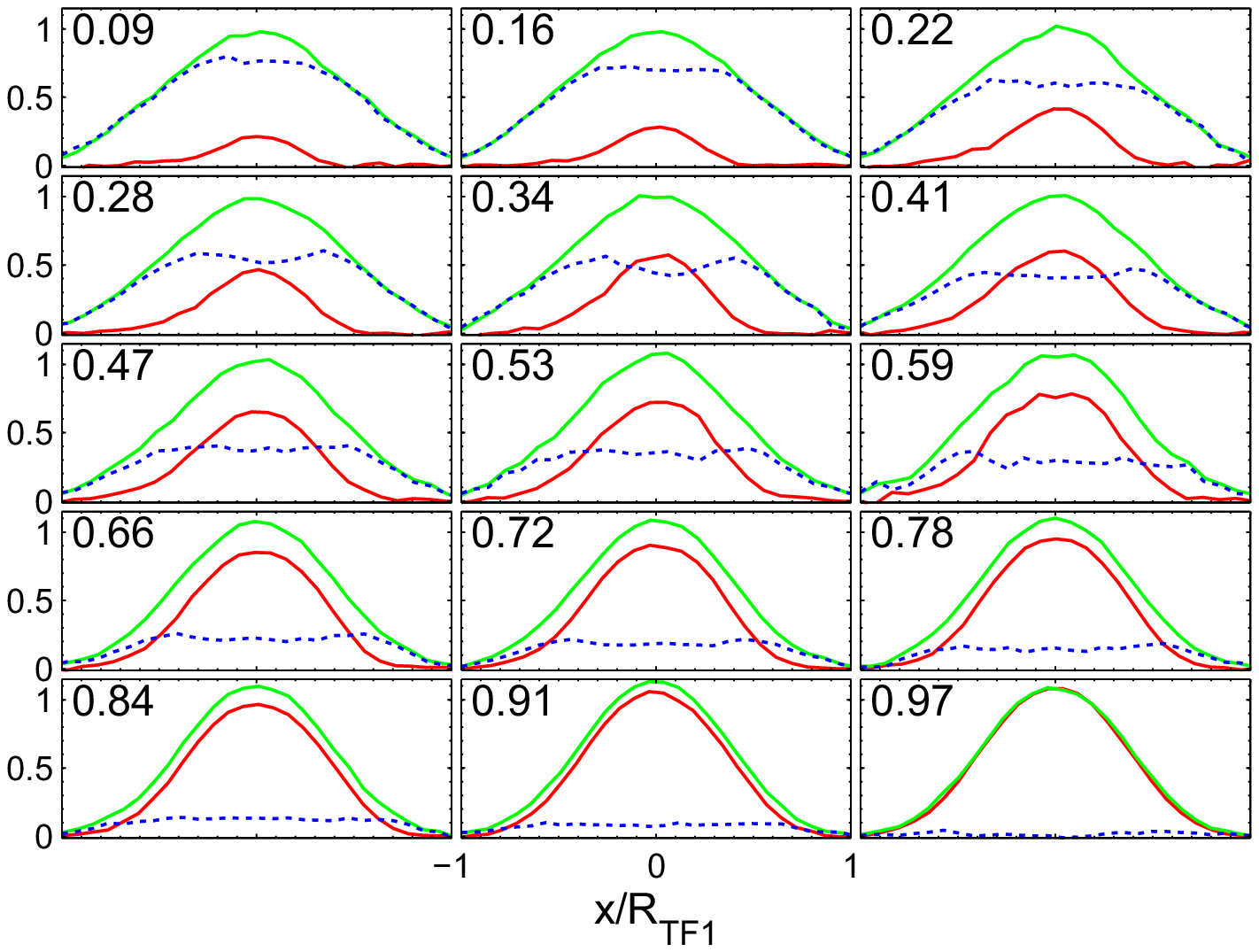}
\end{center}
\caption{Measured column density profiles in units of $N_\uparrow/R_{\mathrm{TF}\uparrow}$ at 775~G, for $E_{\mathrm{F}\uparrow}/\tilde E_{\mathrm{b}}=0.75$ versus $N_\downarrow/N_\uparrow$ ($R_{\mathrm{TF}\uparrow}\equiv\sqrt{2E_{\mathrm{F}\uparrow}/(m\omega_\perp^2)}$ is the majority Thomas--Fermi radius). Green: $\uparrow$-Majority; Red: $\downarrow$-Minority. Blue-dashed: Column density difference. Each profile is labeled by its $N_\downarrow/N_\uparrow$ range. For the density difference, the flat centre and two peaks at the edges are consistent with a fully paired core of the corresponding 2D density profiles.
Adapted with permission from \cite{ThomasSpinImbalancedQ2D2014} copyrighted by the American Physical Society; notations altered.
} \label{fig:PhaseSeparation}
\end{figure}

The phase separation in a quasi-2D gas has been observed by J.~E.~Thomas and co-workers~\cite{ThomasSpinImbalancedQ2D2014}. The images of atoms in the two spin-states are shown in figure~\ref{fig:PhaseSeparation}.
The images are taken along the plane of the motion, therefore, the density profiles are the result of the integration along one dimension. The flat center in the column density difference (blue dashed lines) is a signature of fully balanced core. Qualitatively, therefore, the balanced phase appears in the center similar to that in 3D.

Further studies of the phase separation~\cite{SpinImbalanced2D2016} have shown that there are no discontinuities in the polarization or density profiles at the transition. This is qualitatively different from the 3D phase separation~\cite{KetterleImbalPhaseDiagram2008} where the transition is of the first order. It is suggested \cite{SpinImbalanced2D2016} that the absence of discontinuities in 2D is due to the enhanced role of fluctuations.

The 3rd dimension may be important for the imbalanced 2D gases. A minority-spin atom moving in the field of the majority atoms may have higher lying axial harmonic-oscillator states admixed to its state of motion.

\begin{figure*}[htb!]
\begin{center}
\includegraphics[width=0.8\linewidth]{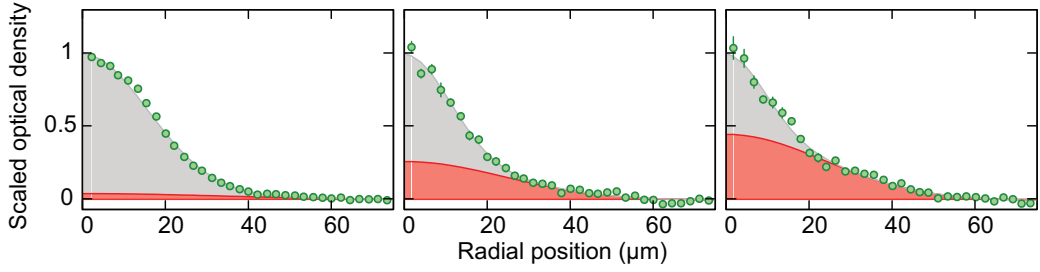}
\end{center}
\caption{Optical density of the minority cloud after 3 ms time of flight normalized to its peak value, with a double Gaussian fit to the data; for polarization $(N_\uparrow-N_\downarrow)/(N_\uparrow+N_\downarrow)=0.25$, 0.55, and 0.75 respectively taken at a field of 780~G. The thermal component is shaded in red, while the condensate is shaded in gray. Error bars represent the standard deviation of the mean in evaluating the azimuthal average. All distributions represent an average of 30 experimental realizations. From \cite{SpinImbalanced2D2016}.} \label{fig:ImbalCondensation}
\end{figure*}

Expansion of the imbalanced gas into the free 3D space upon the instantaneous trap extinction shows a clear bimodal structure in the plane of the initial 2D motion~\cite{SpinImbalanced2D2016}. Such bimodal profiles are shown in figure~\ref{fig:ImbalCondensation}.
The bimodal structure is clearly visible on the Bose side of the 3D Feshbach resonance, where two atoms may remain bound upon removal of the tight confinement. For dimer molecules, the density profile in the 2D plane after long expansion time may be closely related to the velocity profile before the release~\cite{SpinImbalanced2D2016}. This bimodal structure is interpreted as Bose condensation of dimer molecules~\cite{SpinImbalanced2D2016}. Interestingly, the bimodal structure of the expanded gas is visible both in the phase-separation regime as well as outside of this regime. In a weakly-repulsive atomic Bose gas, bimodal structure has been observed without condensation~\cite{Phillips2DBosePhases2009}.

While the week trapping potential in the $xy$ plane is a driver of the phase-separation, the tendency of forming the population-balanced core hides other interesting phenomena, such as the Fulde--Ferell--Larkin--Ovchinnikov superfluidity and the \textit{p}-wave pairing driven by the Kohn--Luttinger mechanism~\cite{KohnLuttingerPPairing1965,FayPPairing1968,MKaganPPairing1988}. Such effects may be seen in a system with a uniform trapping of the atoms in the $xy$ plane.

\subsection{Pressure and polaron effects in the 2D imbalanced case}

If we consider the imbalanced Fermi gas with $n_\uparrow>n_\downarrow$, then the total energy density (energy per unit surface) reads in the BCS domain
\begin{equation}
E=\frac{n_\uparrow\varepsilon_{\mathrm{F}\uparrow}}2+ \frac{n_\downarrow\varepsilon_{\mathrm{F}\downarrow}}2- n_\downarrow|E_{\mathrm{b}}|.
\label{eq:EnDensityBCS}
\end{equation}
Correspondingly the pressure in this case at low temperatures is given by
\begin{equation}
P=n_\uparrow\varepsilon_{\mathrm{F}\uparrow}+n_\downarrow\varepsilon_{\mathrm{F}\downarrow}-f(E),
\end{equation}
where $\varepsilon_{\mathrm{F}\uparrow}\equiv \hbar^2k_{\mathrm{F}\uparrow}^2/(2m_\uparrow)$, $\varepsilon_{\mathrm{F}\downarrow}\equiv \hbar^2k_{\mathrm{F}\downarrow}^2/(2m_\downarrow)$ are the Fermi energies of the ``up'' and ``down'' spins and the function $f(E)$ depends upon the total energy. It is reasonable to assume that the bare ``up'' and ``down'' masses $m_\uparrow=m_\downarrow$ are equal. However the effective ``up'' and ``down'' masses can be different due to many-body fermionic polaron effects~\cite{MKaganPolarons2011,MKaganPolarons2017}.
In the BCS domain at low temperatures:
\begin{equation}
f(E)\simeq\frac12(n_\uparrow\varepsilon_{\mathrm{F}\uparrow}+n_\downarrow\varepsilon_{\mathrm{F}\downarrow}),
\end{equation}
and hence the pressure is approximately
\begin{equation}
P\simeq f(E).
\end{equation}

An interpolating model has been developed in a form applicable both in the Fermi and Bose domain as well as in the intermediate region of strong interactions~\cite{ThomasRFQuasi2D2012,ThomasSpinImbalancedQ2D2014}. In these papers, the energy density is taken in the form
\begin{equation}
E=\frac{n_\uparrow\varepsilon_{\mathrm{F}\uparrow}}2+ \frac{n_\downarrow\varepsilon_{\mathrm{F}\downarrow}}2+ n_\downarrow E_{\mathrm{p}}(\downarrow),
\end{equation}
where $E_{\mathrm{p}}(\downarrow)$ is a negative interaction energy per polaron:
\begin{equation}
E_{\mathrm{p}}(\downarrow)\equiv y_m\left(\frac{\varepsilon_{\mathrm{F}\uparrow}}{|E_{\mathrm{b}}|}\right) \varepsilon_{\mathrm{F}\uparrow}.\label{eq:EpDown}
\end{equation}
In (\ref{eq:EpDown}), the following analytic approximation for the function $y_m$ is used  \cite{Klawunn2DPolaron2011}:
\begin{equation}
y_m\left(\frac{\varepsilon_{\mathrm{F}\uparrow}}{|E_{\mathrm{b}}|}\right)= -\frac2{\ln\left(1+2\frac{\varepsilon_{\mathrm{F}\uparrow}}{|E_{\mathrm{b}}|}\right)}.
\end{equation}
Deep in the BEC domain $E_{\mathrm{p}}(\downarrow)\simeq-|E_{\mathrm{b}}|$ in agreement with (\ref{eq:EnDensityBCS}).

The total pressure of the two spin components may be expressed according to \cite{ThomasRFQuasi2D2012,ThomasSpinImbalancedQ2D2014} as
\begin{equation}
P=n_\uparrow\mu_\uparrow+n_\downarrow\mu_\downarrow-E,
\end{equation}
where $\mu_\uparrow=\partial E/\partial n_\uparrow$ and $\mu_\downarrow=\partial E/\partial n_\downarrow$ are the partial chemical potentials of the respective spin components.

It is important to select the correct geometry of the quasi-2D trap to avoid phase separation which is always present in 3D. Note that in the imbalanced case the phase separation is connected with the negative partial compressibilities (negative sound velocities squared for ``up'' and ``down'' spins, or respectively with the negative values of the derivatives $\partial\mu_\uparrow/\partial n_\uparrow$, $\partial\mu_\downarrow/\partial n_\downarrow$) \cite{MKaganPhaseSeparation2001,MKaganPolarons2011,MKaganPolarons2017}.

\section{Summary and unresolved problems}\label{sec:Concl}

Fermi-to-Bose crossover in 2D ultracold gases has been discussed. While the interactions at close distances are governed by 3D potentials, the many-body systems can be parametrized in terms of parameters conventional for pure 2D description of many-body systems. A number of quantitative measurements are in agreement with 2D models, nevertheless, criteria of two-dimensionality are still under discussion.

There are a lot of open questions (both theoretical and experimental) in the subjects ranging from the experimental observation of the Kohn-Luttinger superfluidity~\cite{KohnLuttingerPPairing1965,FayPPairing1968,MKaganPPairing1988} in strongly-imbalanced 2D Fermi gases to the Fulde--Ferell--Larkin--Ovchinnikov phases~\cite{FuldeFerrell,LarkinOvchinnikov} in case of small imbalance and so on.
Note that a recent experimental discovery of the \textit{p}-wave superconductivity induced by proximity effect in single-layer graphene (on electron-doped oxide superconductor) \cite{PWaveGraphene2017} is building a bridge between an interesting physics of Dirac semimetals and hexagonal 2D optical lattices with the emerging Dirac points~\cite{EsslingerDiracPoints2012}.

\ack
The authors are thankful to Igor Boettcher, Selim Jochim, and Puneet Murthy for valuable discussions.
A.~T. acknowledges the financial support by the program of the Presidium of Russian Academy of Sciences ``Fundamental problems of nonlinear dynamics'' and Russian Foundation for Basic Research (grants No. 14-22-02080-ofi-m, 15-02-08464, 15-42-02638). M.~Yu.~K. greatly acknowledges support from the Basic Research Program of the National Research University Higher School of Economics and RFBR grant 17-02-00135-a.

\end{document}